\documentclass[aps,prx,reprint,superscriptaddress,amsmath,amssymb,floatfix]{revtex4-2}

\usepackage{graphicx}
\usepackage{bm}
\usepackage{xcolor}
\usepackage{tikz}
\usetikzlibrary{arrows.meta,positioning,calc}
\definecolor{archnavy}{HTML}{0F2744}
\definecolor{archteal}{HTML}{0F766E}
\definecolor{archgold}{HTML}{C4A35A}
\definecolor{archmist}{HTML}{F4F7FA}
\definecolor{archslate}{HTML}{64748B}
\usepackage[colorlinks=true,linkcolor=blue!60!black,citecolor=blue!60!black,urlcolor=blue!60!black]{hyperref}

\newif\ifdraft
\drafttrue
\newcommand{\PENDING}[1]{\ifdraft\par\noindent{\color{red!70!black}\textbf{[Pending]} #1}\par\fi}

\graphicspath{{figures/}}

\begin{document}
\raggedbottom

\title{EpNet-Si: Atomic Form Factors Learned from Band Structures}

\author{Qiang Zhu}
\email{qzhu8@charlotte.edu}
\affiliation{Department of Mechanical Engineering and Engineering Science, University of North Carolina at Charlotte, Charlotte, North Carolina 28223, USA}

\date{\today}

\begin{abstract}
The empirical pseudopotential method (EPM) computes band structures with one diagonalization per $k$ point and no self-consistency, but its fixed atomic form factors do not transfer between crystal structures.
To address this limitation, we introduce EpNet-Si, an EPM for silicon in which an equivariant network predicts each atom's plane-wave form factor from its local environment: a species potential plus monopole, dipole and quadrupole corrections, mixed by a learned router over atomic densities, together with an $s/p$ nonlocal projector.
With 6{,}195 parameters, the model is trained through the eigensolver on reference density functional theory (DFT) band energies of only 77 cells.
Among 231 distortions of the trained structure types, the 142 gapped cells have mean band and gap errors of 15 and 8\,meV; among 192 unseen structures, the 165 gapped cells have mean errors of 38 and 49\,meV.
Physical properties derived from EpNet-Si, including effective masses, gap pressure coefficients, deformation potentials and the density of states at the Fermi level, agree with those obtained from the reference DFT calculations.
These results suggest that the classical EPM, fitted to a small set of reference data, can be extended into a fast alternative for modern electronic band-structure calculations.
\end{abstract}

\maketitle

\section{Introduction}
\label{sec:intro}

The empirical pseudopotential method (EPM) was the first practical route to the band structures of real solids.
Building on the observation of Phillips, Kleinman~\cite{Phillips1959}, Cohen and Heine~\cite{Cohen1961} that valence electrons feel only a weak, smooth effective potential, Cohen and Bergstresser showed in 1966 that three Fourier coefficients per element reproduce the optical spectra of fourteen diamond and zinc-blende semiconductors~\cite{Cohen1966}.
Nonlocal, angular-momentum-dependent terms and spin--orbit coupling were added later~\cite{Chelikowsky1976,CohenChelikowsky1988}, and the semi-empirical pseudopotentials of Wang and Zunger, fitted to local-density calculations and combined with the folded-spectrum method, extended the approach to nanocrystals of $10^4$--$10^6$ atoms~\cite{Wang1994,Wang1995}.
The EPM Hamiltonian $H=T+V_{\rm ps}$, with $V_{\rm ps}$ a superposition of fixed atomic form factors $v(|\mathbf G|)$, requires one diagonalization per $k$ point in a small plane-wave basis, without self-consistency.

Two methods have since taken over the roles the EPM once played.
Kohn--Sham density-functional theory (DFT)~\cite{Hohenberg1964,Kohn1965} with \emph{ab initio} norm-conserving, ultrasoft or projector-augmented-wave pseudopotentials~\cite{Blochl1994,Kresse1999} is transferable: the same pseudopotential serves any structure without fitting, at the price of a self-consistent field (SCF) cycle, dense $k$ grids and a plane-wave basis roughly ten times larger than an EPM needs.
Empirical tight binding~\cite{Slater1954,Vogl1983,Jancu1998} has the smallest cost, a few orbitals per atom and sparse matrices, and is the standard for device and transport simulations, but its parameters transfer poorly across bonding types.
The limited transferability of the classical EPM reflects its use of a fixed, spherical atomic $v(|q|)$ that is independent of the local environment.
Combining the low computational cost of an EPM with transferability across diverse bonding environments remains a challenge.

Machine learning has recently changed several of the constraints that fixed the classical EPM.
Graph neural networks can make each atom's form factor a function of its neighborhood; $E(3)$-equivariant networks~\cite{Batzner2022} can add angular ($l\le2$) components with the correct transformation law by construction. In addition, automatic differentiation through a Hermitian eigensolver allows band energies, not potentials, to serve as the training target. 
Related efforts have taken parts of this path.
Kim and Son learned environment-dependent atomic potentials whose sum reproduces the local part of the converged Kohn--Sham potential, $V_{\rm loc}+V_H+V_{xc}$, using rotation-covariant density and SOAP descriptors~\cite{KimSon2024}, while keeping the nonlocal projectors of the DFT pseudopotential and its plane-wave cutoff.
DeepPseudopot learns a neural radial local potential with nonlocal and spin--orbit terms from $GW$ band structures of zinc-blende and diamond semiconductors~\cite{Lin2025}.
Separately, other works on DeepH, HamGNN and DeePTB~\cite{Li2022DeepH,Zhong2023HamGNN,Gu2024DeePTB} and their long-range extensions~\cite{Zhong2026LR,Zhang2026Ewald} predict the full DFT Hamiltonian in an atomic-orbital basis with large models, and their training labels are converged SCF Hamiltonians.

To our knowledge, no model has learned atom-resolved, equivariant plane-wave form factors end to end from band energies and then been tested for transfer across qualitatively different bonding environments.
This is the gap we address here, and the choice of material follows from it.
For an empirical pseudopotential, the hard test is not a new element but a new environment, and silicon supplies many environments with a single species: one element forms fourfold diamond and its hexagonal polytypes~\cite{Hauge2015,Fadaly2020}, guest-free open-cage clathrates with wide gaps~\cite{Gryko2000,Kim2015Si24}, dense BC8 and R8 phases and sixfold to eightfold metals under pressure~\cite{Mujica2003}, and thousands of predicted $sp^3$ frameworks.

In this work, we introduce EpNet-Si, a neural empirical pseudopotential for silicon.
Given only lattice vectors and atomic positions, an equivariant network predicts each atom's form factor as a smooth species potential plus an environment-dependent correction; the result is assembled into a plane-wave Hamiltonian $H(\mathbf k)$ with a small $s/p$ nonlocal term and diagonalized once per $k$ point.
We train the model on the Kohn--Sham PBE~\cite{Perdew1996} band energies of 77 cells and test it on 423 structures in two groups: distortions of the trained structure types, and structure types it has never seen.
EpNet-Si reproduces the distortions at training accuracy with no failures.
On unseen structures it preserves the PBE band shapes, and its errors are conduction-band offsets that are largest in the most open frameworks.
Leave-one-family-out runs show which training data matter, an ablation and the learned router show what the network relies on, and effective masses, gap pressure coefficients, deformation potentials and the density of states at the Fermi level transfer to structures the model never saw.
We close by discussing what these experiments reveal about the limits of a local form factor, and how EpNet-Si can be extended to other elements, compounds and reference theories.

\section{Methods}
\label{sec:methods}

\subsection{EPM applied to diamond Si}
\label{sec:epm}

\begin{figure}
\includegraphics[width=0.9\columnwidth]{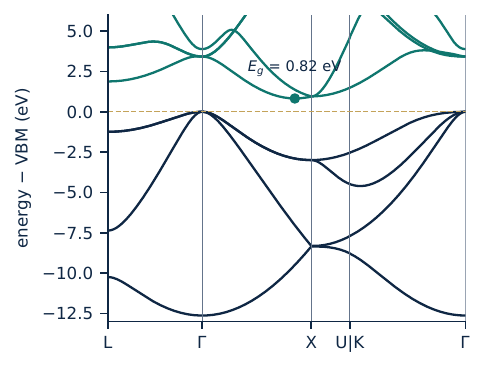}
\caption{\label{fig:cohen}Band structure of diamond Si from the three Cohen--Bergstresser form factors $V_3$, $V_8$, $V_{11}$~\cite{Cohen1966}, in a basis of 113 plane waves ($|\mathbf G|^2\le20\,(2\pi/a)^2$). Valence bands navy, conduction bands teal; the dot marks the conduction-band minimum at $0.85\,X$ (indirect gap 0.82\,eV).}
\end{figure}

In the plane-wave basis $\{|\mathbf k+\mathbf G\rangle\}$ the EPM Hamiltonian of a cell of volume $\Omega$ with atoms at $\bm\tau_a$ is
\begin{align}
H_{\mathbf G\mathbf G'}(\mathbf k)={}&\frac{\hbar^2}{2m}|\mathbf k+\mathbf G|^2\delta_{\mathbf G\mathbf G'}
+\frac{1}{\Omega}\sum_a v_a(\Delta\mathbf G)\,e^{-i\Delta\mathbf G\cdot\bm\tau_a}\nonumber\\
&+\Delta H^{\rm nl}_{\mathbf G\mathbf G'}(\mathbf k),
\label{eq:H}
\end{align}
with $\Delta\mathbf G=\mathbf G-\mathbf G'$.
In the classical EPM $v_a=v(|\Delta\mathbf G|)$ is one fixed function per species and $\Delta H^{\rm nl}$ is absent or a fixed angular-momentum-dependent correction~\cite{Cohen1966,Chelikowsky1976}.
For diamond Si, Eq.~\eqref{eq:H} reduces to the Cohen--Bergstresser form $V_S(|\Delta\mathbf G|)\cos(\Delta\mathbf G\cdot\bm\tau)$ with $\bm\tau=\tfrac a8(1,1,1)$~\cite{Cohen1966}.
The reciprocal lattice vectors are $\mathbf G=\tfrac{2\pi}{a}(h,k,l)$ with $h,k,l$ all even or all odd, and the form factors are labeled by $|\mathbf G|^2$ in units of $(2\pi/a)^2$.
Only three are nonzero, $V_3=-0.21$, $V_8=+0.04$ and $V_{11}=+0.08$\,Ry on the $\langle111\rangle$, $\langle220\rangle$ and $\langle311\rangle$ shells.
The structure factor $\cos(\mathbf G\cdot\bm\tau)=\cos[\tfrac\pi4(h+k+l)]$ vanishes on the $\langle200\rangle$, $\langle222\rangle$ and $\langle420\rangle$ shells ($h+k+l\equiv2 \bmod 4$); $V_0$ only shifts all energies; and the weak, smooth pseudopotential is taken as zero beyond $|\mathbf G|^2=11$, which removes $\langle400\rangle$ and $\langle331\rangle$.
The basis contains every $\mathbf G$ with $|\mathbf G|^2\le20$, i.e.\ the first nine shells ($1+8+6+12+24+8+6+24+24=113$ plane waves, through $\langle420\rangle$).
Fig.~\ref{fig:cohen} shows the resulting band structure: an indirect gap with the conduction-band minimum at $0.85\,X$ and a direct gap at $\Gamma$ of 3.42\,eV.

\subsection{EPM with environment-dependent form factors}
\label{sec:model}

Unlike the highly symmetric diamond case, the reciprocal lattice in a general crystal samples $v(|q|)$ at many more values of $|\mathbf G|$ below the cutoff of the species potential ($q_c=8$\,\AA$^{-1}$, Eq.~\eqref{eq:sphere}), and the number of contributing shells grows with the number of atoms in the primitive cell, so the whole function $v(|q|)$ is needed rather than a few numbers.
Even with a smooth $v(|q|)$ defined at every $q$, it is problematic to assume that a Si atom scatters electrons in the same way, whether it sits on a tetrahedral site, in an open cage or in a dense metal.
Moreover, once distortions break the local site symmetry, the potential acquires an angular dependence that a spherical $v(|q|)$ cannot express.

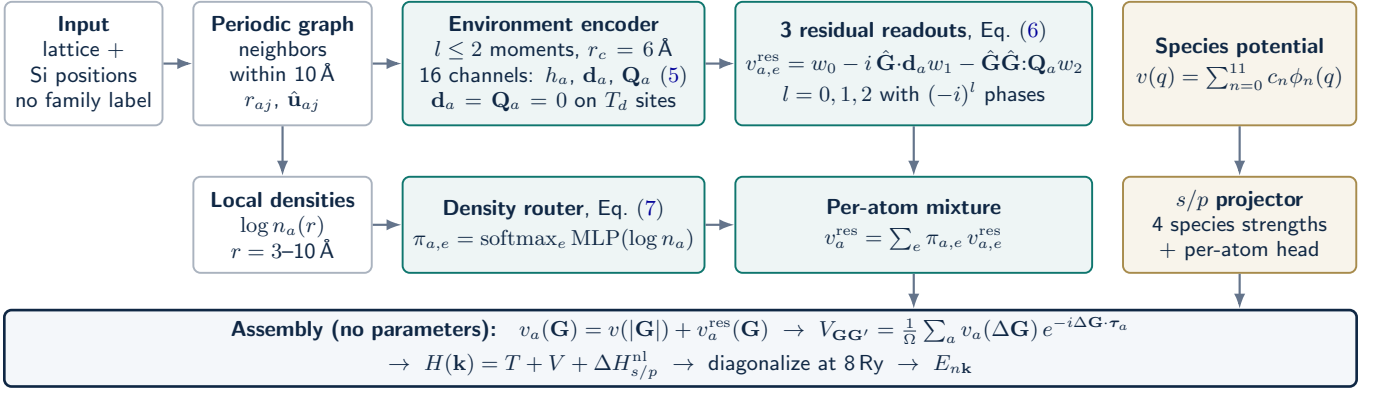
\begin{figure*}
\resizebox{\textwidth}{!}{%
\begin{tikzpicture}[
  font=\sffamily\small,
  box/.style={rounded corners=3pt,inner sep=3pt,align=center,anchor=center,line width=0.8pt,text=archnavy},
  fixed/.style={box,draw=archslate!55,fill=white},
  learn/.style={box,draw=archteal,fill=archteal!7},
  spec/.style={box,draw=archgold!85!black,fill=archgold!12},
  dark/.style={box,draw=archnavy,fill=archmist,line width=1.1pt},
  arr/.style={-{Latex[length=2.2mm]},line width=0.9pt,archslate},
  c1/.style={minimum width=2.3cm,text width=2.0cm},
  c2/.style={minimum width=2.6cm,text width=2.3cm},
  c3/.style={minimum width=4.4cm,text width=4.1cm},
  c4/.style={minimum width=5.2cm,text width=4.9cm},
  c5/.style={minimum width=3.4cm,text width=3.1cm},
  r1/.style={minimum height=1.8cm},
  r2/.style={minimum height=1.4cm},
]
\def\xa{1.15}\def\xb{4.05}\def\xc{8.00}\def\xd{13.25}\def\xe{18.00}
\def\ya{0}\def\yb{-2.35}\def\yc{-4.15}

\node[fixed,c1,r1] (IN)  at (\xa,\ya) {{\bfseries Input}\\ lattice +\\ Si positions\\ no family label};
\node[fixed,c2,r1] (GR)  at (\xb,\ya) {{\bfseries Periodic graph}\\ neighbors\\ within 10\,\AA\\ $r_{aj}$, $\hat{\mathbf u}_{aj}$};
\node[learn,c3,r1] (ENC) at (\xc,\ya) {{\bfseries Environment encoder}\\ \mbox{$l\le2$} moments, $r_c=6$\,\AA\\ \mbox{16 channels: $h_a$, $\mathbf d_a$, $\mathbf Q_a$ \eqref{eq:enc}}\\ $\mathbf d_a=\mathbf Q_a=0$ on $T_d$ sites};
\node[learn,c4,r1] (RO)  at (\xd,\ya) {{\bfseries 3 residual readouts}, Eq.~\eqref{eq:readout}\\[2pt] \mbox{$v^{\rm res}_{a,e}=w_0-i\,\hat{\mathbf G}{\cdot}\mathbf d_a w_1-\hat{\mathbf G}\hat{\mathbf G}{:}\mathbf Q_a w_2$}\\[1pt] \mbox{$l=0,1,2$ with $(-i)^l$ phases}};
\node[spec,c5,r1]  (SPH) at (\xe,\ya) {{\bfseries Species potential}\\[2pt] \mbox{$v(q)=\sum_{n=0}^{11}c_n\phi_n(q)$}};

\node[fixed,c2,r2] (DEN) at (\xb,\yb) {{\bfseries Local densities}\\ $\log n_a(r)$\\ \mbox{$r=3$--10\,\AA}};
\node[learn,c3,r2] (RT)  at (\xc,\yb) {{\bfseries Density router}, Eq.~\eqref{eq:router}\\[2pt] \mbox{$\pi_{a,e}=\mathrm{softmax}_e\,\mathrm{MLP}(\log n_a)$}};
\node[learn,c4,r2] (MIX) at (\xd,\yb) {{\bfseries Per-atom mixture}\\[2pt] \mbox{$v^{\rm res}_a=\sum_{e}\pi_{a,e}\,v^{\rm res}_{a,e}$}};
\node[spec,c5,r2]  (SP)  at (\xe,\yb) {{\bfseries $s/p$ projector}\\ 4 species strengths\\ + per-atom head};

\node[dark,minimum width=19.7cm,text width=19.5cm,minimum height=1.1cm] (ASM) at (9.85,\yc) {%
  \mbox{{\bfseries Assembly (no parameters):}\quad
  $v_a(\mathbf G)=v(|\mathbf G|)+v^{\rm res}_a(\mathbf G)$
  \;$\to$\; $V_{\mathbf{GG}'}=\tfrac1\Omega\sum_a v_a(\Delta\mathbf G)\,e^{-i\Delta\mathbf G\cdot\bm\tau_a}$}\\[2pt]
  \mbox{$\to$\; $H(\mathbf k)=T+V+\Delta H^{\rm nl}_{s/p}$
  \;$\to$\; diagonalize at 8\,Ry \;$\to$\; $E_{n\mathbf k}$}};

\draw[arr] (IN) -- (GR);
\draw[arr] (GR) -- (ENC);
\draw[arr] (ENC) -- (RO);
\draw[arr] (DEN) -- (RT);
\draw[arr] (RT) -- (MIX);
\draw[arr] (GR) -- (DEN);
\draw[arr] (RO) -- (MIX);
\draw[arr] (SPH) -- (SP);
\draw[arr] (MIX) -- (MIX |- ASM.north);
\draw[arr] (SP) -- (SP |- ASM.north);
\end{tikzpicture}}
\caption{\label{fig:arch}Architecture of EpNet-Si. Fixed operations (grey frames), components learned per atom (teal) and per species (gold); the bottom frame is the parameter-free assembly. The environment encoder produces invariant ($h_a$), vector ($\mathbf d_a$) and rank-2 ($\mathbf Q_a$) features; three residual readouts are mixed per atom by a router over local densities and added to the species potential; the result is assembled with an $s/p$ nonlocal term into $H(\mathbf k)$, Eq.~\eqref{eq:H}, and diagonalized.}
\end{figure*}

When designing EpNet-Si, we aim to keep the Hamiltonian of Eq.~\eqref{eq:H} and make each atomic form factor a function of its local environment.
We require this factor to reduce to a smooth species potential when the environment correction vanishes, to transform correctly under rotations and inversion for any arrangement of atoms, and to vary continuously with the environment. Taking all factors into account, we end up with the model illustrated in Fig.~\ref{fig:arch}.

First, we express the form factor of atom $a$ ($v_a$) as a mixture of a species potential and three environment-dependent residual readouts in the spirit of mixture-of-experts models~\cite{Jacobs1991,Shazeer2017}:
\begin{equation}
v_a(\mathbf G)=v(|\mathbf G|)+\sum_{e=1}^{3}\pi_{a,e}\,v^{\rm res}_{a,e}(\mathbf G),
\label{eq:va}
\end{equation}
where $e=1,2,3$ labels the readout (an ``expert'' in mixture-of-experts terminology) and the per-atom weights $\pi_{a,e}\ge0$, with $\sum_e\pi_{a,e}=1$, are set by a router that blends all three readouts continuously.
Removing the residual form factors and per-atom $s/p$ corrections recovers an EPM with $v_a=v$ and fixed nonlocal projectors; setting the species $s/p$ strengths to zero then gives the purely local limit.

Second, the \emph{species potential} is expanded as
\begin{equation}\label{eq:sphere}
v(q)=\tfrac12(1+\cos\pi x)\sum_{n=0}^{11}c_n\cos n\pi x,
\end{equation}
with $x=\min(q/q_c,1)$ and $q_c=8$\,\AA$^{-1}$, so that $v$ is smooth and vanishes beyond $q_c$.
The coefficients $c_n$ were initialized from the Cohen--Bergstresser form factors of Sec.~\ref{sec:epm}.
We converted these to per-atom form factors using $v=V_S\Omega_{\rm cell}/2$, with $\Omega_{\rm cell}=a^3/4$ and $a=5.431$\,\AA.
We joined the three converted values by a raised-cosine interpolation, held $v$ at its $\langle111\rangle$ value below 2.0\,\AA$^{-1}$, brought it to zero at 6.5\,\AA$^{-1}$, and fitted the $c_n$ to this curve by least squares on 81 points between 0 and $q_c$.
They were then refined during model development, and EpNet-Si was initialized from those values.
The $G=0$ component is set to zero, which fixes the energy reference.

Third, each atom's environment within $r_c=6$\,\AA\ is encoded, under periodic boundary conditions, by neighbor moments up to $l=2$ written in Cartesian form.
With a radial basis $B_n(r)=\mathrm{sinc}(n\pi r/r_c)\,f(r/r_c)$, $n=1,\dots,8$, where $\mathrm{sinc}(x)=\sin x/x$ ($\mathrm{sinc}(0)=1$) and $f(x)=1-10x^3+15x^4-6x^5$ is a $C^2$ envelope, and a learned linear map to 16 channels, the edge weights are $\mathbf w_{aj}=(W_{\rm rad}\mathbf B(r_{aj}))\odot\mathbf e_{s_j}$ with species embedding $\mathbf e$, central atom $a$ and neighbor $j$.
The encoder works in a single pass. It first forms the ungated neighbor moments
\begin{align}
\mathbf s_a&=\mathbf e_{s_a}+\textstyle\sum_j\mathbf w_{aj},\qquad
\mathbf D_a=\textstyle\sum_j\mathbf w_{aj}\,\hat{\mathbf u}_{aj},\nonumber\\
(\mathbf T_a)_{\alpha\beta}&=\textstyle\sum_j\mathbf w_{aj}\big(\hat u_{aj,\alpha}\hat u_{aj,\beta}-\tfrac13\delta_{\alpha\beta}\big),\label{eq:moments}
\end{align}
where $\hat{\mathbf u}_{aj}$ is the bond direction from atom $a$ to neighbor $j$, $\alpha,\beta\in\{x,y,z\}$, and subtracting $\tfrac13\delta_{\alpha\beta}$ makes $\mathbf T_a$ traceless, i.e.\ a pure $l=2$ quantity.
A multilayer perceptron (MLP) with one hidden layer of 32 units and the sigmoid linear unit $\mathrm{SiLU}(x)=x\,\sigma(x)$ as activation, where $\sigma(x)=1/(1+e^{-x})$ is the logistic sigmoid, maps the 48 invariants of these moments, $[\mathbf s_a,|\mathbf D_a|,|\mathbf T_a|]$, to three gates $\mathbf g_{0,1,2}$ of 16 channels each.
The norms are taken separately for each channel: the Euclidean norm for $\mathbf D_a$ and the Frobenius norm for $\mathbf T_a$.
The gates are applied once to give the per-atom features
\begin{align}
h_a&=\mathbf s_a+\mathbf g_0,\qquad
\mathbf d_a=\sigma(\mathbf g_1)\odot\mathbf D_a,\nonumber\\
\mathbf Q_a&=\sigma(\mathbf g_2)\odot\mathbf T_a.
\label{eq:enc}
\end{align}
Because the gates depend only on invariants, $h_a$ is invariant, $\mathbf d_a$ rotates as a vector and $\mathbf Q_a$ as a traceless symmetric tensor.

Fourth, each readout $e$ maps the encoder features to a form factor with $l=0,1,2$ components,
\begin{align}
v^{\rm res}_{a,e}(\mathbf G)={}&\sum_r (W^{(e)}_0h_a+b^{(e)})_r R_r(G)\nonumber\\
&+\sum_n (U^{(e)}h_a+c^{(e)})_n C_n(G)\nonumber\\
&-i\sum_r (W^{(e)}_1\mathbf d_a)_r\cdot\hat{\mathbf G}\,R_r(G)\nonumber\\
&-\sum_r \hat{\mathbf G}\cdot(W^{(e)}_2\mathbf Q_a)_r\cdot\hat{\mathbf G}\,R_r(G),
\label{eq:readout}
\end{align}
where $R_r$ ($r=1,\dots,8$) is a sinc-envelope basis in reciprocal space multiplied by the raised-cosine envelope of Eq.~\eqref{eq:sphere}, and $C_n$ ($n=0,\dots,11$) is the cosine basis of Eq.~\eqref{eq:sphere}, so that every term of $v_a$, like the species potential, vanishes smoothly at $q_c=8$\,\AA$^{-1}$.
The prefactors $+1$, $-i$ and $-1$ of the $l=0$, 1 and 2 terms are $(-i)^l$: they make the dipole term odd in $\mathbf G$ and imaginary, and the monopole and quadrupole terms even and real, so that together with the structure factor in Eq.~\eqref{eq:H} the potential satisfies $V(-\mathbf G)=V(\mathbf G)^*$ for any atomic arrangement.
Rotating the structure rotates $\mathbf d_a$, $\mathbf Q_a$ and $\hat{\mathbf G}$ together, leaving the spectrum unchanged.

Finally, the mixture weights in Eq.~\eqref{eq:va} are given by the density router,
\begin{equation}
\pi_{a,e}=\frac{e^{o_{a,e}}}{\sum_{e'}e^{o_{a,e'}}},\qquad \mathbf o_a=\mathrm{MLP}(\mathbf z_a),
\label{eq:router}
\end{equation}
where $\mathbf z_a$ collects the logarithms of the local atomic density $n_a(r_k)$ at six radii $r_k=3$, 4, 5, 6, 8 and 10\,\AA, each standardized with fixed constants (a family-balanced mean and variance computed once over the training atoms of an earlier, larger training set).
Here $n_a(r_k)$ is a smooth count of the neighbors within $r_k$ (weight one up to $0.8r_k$, falling to zero at $r_k$ with a raised cosine), divided by the sphere volume $\tfrac43\pi r_k^3$ and by a reference density of 0.05\,\AA$^{-3}$.
The radii span the first neighbor shell to the medium-range packing that separates open, fourfold and metallic Si, and the logarithm compresses the wide range of densities between them.
The MLP has one hidden layer of 16 units and three outputs.

\subsection{Nonlocal $s/p$ projector}
\label{sec:nl}

A local $v(\mathbf G-\mathbf G')$ cannot distinguish $s$-like from $p$-like combinations of plane waves.
We add a separable Kleinman--Bylander-type term~\cite{Kleinman1982},
\begin{align}
\Delta H^{\rm nl}_{\mathbf G\mathbf G'}(\mathbf k)=\sum_{a,r}\Big[&D^s_{a,r}\,\beta^s_{ar}(\mathbf p)\beta^{s*}_{ar}(\mathbf p')\nonumber\\
&+D^p_{a,r}\sum_{j}\beta^{p_j}_{ar}(\mathbf p)\beta^{p_j*}_{ar}(\mathbf p')\Big],
\label{eq:nl}
\end{align}
with $\mathbf p=\mathbf k+\mathbf G$, $\mathbf p'=\mathbf k+\mathbf G'$ and Gaussian projectors
\begin{equation}
\beta^s_{ar}(\mathbf p)=\frac{e^{-i\mathbf p\cdot\bm\tau_a}}{\sqrt\Omega}e^{-|\mathbf p|^2/2\sigma_r^2},\qquad
\beta^{p_j}_{ar}(\mathbf p)=\sqrt3\,\frac{p_j}{\sigma_r}\,\beta^s_{ar}(\mathbf p),
\end{equation}
of fixed widths $\sigma_1=1.5$ and $\sigma_2=3.5$\,\AA$^{-1}$.
The strengths $D^{s,p}_{a,r}$ are the sum of four species values and a per-atom correction.
The correction is predicted by a small MLP ($50\to16\to4$, initialized to zero) from the 48 encoder invariants and two coordination numbers, $N_a(3\,\text{\AA})/4$ and $N_a(10\,\text{\AA})/150$, where $N_a(R)$ counts the neighbors of atom $a$ with weight one up to $0.8R$, falling to zero at $R$ with a raised cosine (the count used for the router densities, Sec.~\ref{sec:model}); the divisors bring both inputs to order one.
The projector borrows the Kleinman--Bylander form but has empirical parameters, with no norm-conservation or scattering constraint.
The four species strengths start from zero and are learned with the rest of the model; in the trained model they are $D^s=18.6$ and $6.9$\,eV\,\AA$^3$ and $D^p=8.7$ and $-5.8$\,eV\,\AA$^3$ for $\sigma_1$ and $\sigma_2$.

\subsection{Plane-wave basis and diagonalization}
\label{sec:basis}

EpNet-Si uses a kinetic-energy cutoff of 8\,Ry, i.e.\ $|\mathbf k+\mathbf G|\le G_{\rm max}=5.35$\,\AA$^{-1}$ (about 50 plane waves per atom, roughly one tenth of the number used at the reference cutoff of 40\,Ry).
As in plane-wave DFT codes the basis is centred on each $\mathbf k$, so a crystal described by any supercell has exactly the same bands: the supercell basis at $\mathbf k$ is the union of the primitive-cell bases at the $\mathbf k$ points that fold onto it.
We verified this for diamond (2, 8 and 16 atoms) and clathrate~I (46 and 92 atoms), whose gaps agree to within 0.5\,meV in our single-precision calculations.
A sharp cutoff, however, makes $E(\mathbf k)$ discontinuous, because plane waves enter the basis as $\mathbf k$ moves; at 8\,Ry the jumps reach 25\,meV (13\,meV in the lowest conduction band) and spoil band curvatures.
We therefore add a thin buffer $G_{\rm max}<|\mathbf k+\mathbf G|\le1.05\,G_{\rm max}$ in which the coupling of a plane wave to all others is scaled by $w=\cos^2(\pi x/2)$, with $x=(|\mathbf k+\mathbf G|-G_{\rm max})/(0.05\,G_{\rm max})$, and its kinetic energy is raised by $50x^2$\,eV.
A plane wave thus enters the basis decoupled, $E(\mathbf k)$ is continuous, and every plane wave up to $G_{\rm max}$ remains fully coupled.
Each $\mathbf k$ is folded into the first Brillouin zone before $H(\mathbf k)$ is built, which keeps the basis sphere small.
The Hamiltonian is assembled in single-precision complex arithmetic and diagonalized with a dense Hermitian eigensolver.
In training, gradients are propagated through the lowest $n_b$ eigenvalues, where $n_b$ is the number of bands in the reference.
The model has 6{,}195 parameters in total: the species potential (12), the encoder (3{,}296), three readouts ($3\times612$), the router (163), the $s/p$ species strengths (4) and the per-atom $s/p$ head (884).
Of these, 6{,}147 are trainable; the other 48 are the radial biases of the dipole and quadrupole terms of the three readouts ($3\times2\times8$), which are held at zero because a constant Cartesian bias would select a fixed direction and break rotational covariance.

\subsection{Reference data}
\label{sec:data}

All reference band structures were computed with one workflow in Quantum ESPRESSO 7.6~\cite{Giannozzi2009,Giannozzi2017}.
We used the PBE functional~\cite{Perdew1996} and the \texttt{Si.pbe-n-kjpaw\_psl.1.0.0} PAW dataset~\cite{Blochl1994,DalCorso2014}, with cutoffs of 40\,Ry (wavefunctions) and 320\,Ry (density), no spin polarization and no spin--orbit coupling.
SCF runs used a $4\times4\times4$ Monkhorst--Pack grid, fixed occupations and a convergence threshold of $10^{-10}$\,Ry, and computed $n_{\rm occ}+\max(8,n_{\rm occ}/4)$ bands, or $+\max(16,n_{\rm occ}/4)$ for the test sets generated last.
Band paths follow the SeeK-path~\cite{Hinuma2017} conventions.
Cells whose path gap is at most 0.02\,eV were relabeled as metals so that $E_F$ is physical.
The relabeling used a Marzari--Vanderbilt smeared SCF~\cite{Marzari1999} (0.01\,Ry, $k$ spacing $\le0.15$\,\AA$^{-1}$), path bands, and a uniform-grid non-SCF calculation ($\approx0.12$\,\AA$^{-1}$, no symmetry reduction).
The target is PBE, not experiment; for diamond the target indirect gap is 0.56\,eV.

The training set (Table~\ref{tab:data}) is deliberately small: 77 cells, 48 gapped and 29 metallic.
Twelve catalog families span fourfold semiconductors (diamond, the 4H and lonsdaleite polytypes, ST12, oC12 and bct), the clathrate~I cage, the dense BC8, R8 and tI16 phases, and the sixfold to eightfold metals $\beta$-Sn and simple hexagonal~\cite{Mujica2003}.
Each fully sampled family contributes five cells: the parent, a compressed and an expanded hydrostatic cell ($\mp4\%$ volume), one uniaxial strain and one rattled cell.
Eleven OQMD structures~\cite{Kirklin2015} (eight frameworks, three metals) are included whole.
Two further groups were selected by structural rules that do not depend on any model result: six SACADA-derived frameworks~\cite{Hoffmann2016}, the gapped cell with the fewest atoms in each 0.5\,\AA$^3$/atom bin of volume per atom between 19.8 and 22.8\,\AA$^3$/atom, and three cells whose largest void (3.4--3.7\,\AA, the largest distance from any point to its nearest atom) exceeds that of clathrate~I.
These 66 cells, with only the expanded hydrostatic cell per family, trained a precursor of EpNet-Si.
After that precursor had been tested, we added the compressed hydrostatic cell of the ten families that lacked one, and one open framework, Amsler C07 (28 atoms, 24.0\,\AA$^3$/atom), the gapped Amsler framework with the fewest atoms above 23.5\,\AA$^3$/atom.
The training-set row of Table~\ref{tab:design} summarizes how this set was arrived at.
Every other QE-PBE cell is a test cell.

\begin{table}
\caption{\label{tab:data}The 77 training cells.}
\begin{ruledtabular}
\begin{tabular}{llr}
Structures & Type & Cells \\
\hline
Diamond, 4H, oC12, bct & fourfold & 5 each \\
Lonsdaleite, ST12 & fourfold & 4, 3 \\
Clathrate I & open cage & 5 \\
BC8, R8, tI16 & dense, metallic & 5, 4, 5 \\
$\beta$-Sn, simple hex. & 6--8-fold metal & 5 each \\
OQMD structures & mixed$^{\rm a}$ & 11 \\
Frameworks (SACADA, Amsler) & fourfold nets & 7 \\
Large-void cells & open & 3 \\
\hline
Total & & 77 \\
\end{tabular}
\end{ruledtabular}
\raggedright\footnotesize $^{\rm a}$Eight frameworks and three metals.
\end{table}

\begin{table}
\caption{\label{tab:evalsets}The 423 test structures.}
\begin{ruledtabular}
\setlength{\tabcolsep}{4pt}
\begin{tabular}{@{}lrp{4.3cm}@{}}
Set & Cells & \raggedright Content \tabularnewline
\hline
\multicolumn{3}{l}{\emph{1. Distortions of seen structures} (231)} \\
Single & 147 & \raggedright volume $\pm2$--6\%, strain 1.5--4\%, shear 3--4\%, rattle 0.03\,\AA \tabularnewline
Combined & 84 & \raggedright random 3D strain 2.5--3.5\%, strain + rattle, volume + shear, rattle 0.04\,\AA \tabularnewline
\hline
\multicolumn{3}{l}{\emph{2. Unseen structures} (192)} \\
Clathrate II & 22 & \raggedright parent and distortions; no cage of this type trained \tabularnewline
SACADA frameworks & 97 & \raggedright carbon nets converted to Si, Tersoff-relaxed \tabularnewline
Amsler frameworks & 35 & \raggedright low-density Si frameworks~\cite{Amsler2015} \tabularnewline
OQMD & 17 & \raggedright frameworks and metals outside training~\cite{Kirklin2015} \tabularnewline
AFLOW & 11 & \raggedright structures without a duplicate among our cells~\cite{Curtarolo2012} \tabularnewline
Polytypes etc. & 10 & \raggedright stacking polytypes of 5--15 layers; one random structure \tabularnewline
\end{tabular}
\end{ruledtabular}
\end{table}

Table~\ref{tab:evalsets} lists the 423 test structures (307 gapped, 116 metallic): 231 distortions of the twelve trained structure types and 192 structures of types absent from training.
The 147 single distortions comprise catalog cells excluded from training, validation cells and separately generated strains and rattles; the 84 combined distortions come from two development test sets.
For SACADA, the structures are originally relaxed for carbon, and they were converted to Si and relaxed with the Tersoff potential~\cite{Tersoff1988}; AFLOW structures that duplicate one of our cells (by structure matching and space group) were removed, and all AFLOW cells were relabeled with our workflow rather than using their VASP bands.
Two evaluation sets of 78 and 84 structures were used for model development and training-cell selection.
Their cells are included in the two evaluation groups; we call the 84-structure set, which spans both groups, the development set.
Here, ``test'' denotes cells excluded from the band-energy training loss, not necessarily cells unused in model or training-set selection.
Errors for cells unused in model and training-set selection are not reported separately.
Together with the training set, 500 QE-PBE band structures are used.

\subsection{Loss and training}
\label{sec:loss}

The loss compares model and PBE eigenvalues on the reference $k$ path,
\begin{equation}
\mathcal L=\sum_{f=1}^{N_{\rm fam}}\frac{1}{N_{\rm fam}N_f}\sum_{c\in f}\Big[\ell_{\rm win}(c)+\lambda_g\,\ell_{\rm gap}(c)+\lambda_d\,\ell_{\rm deep}(c)\Big],
\label{eq:loss}
\end{equation}
where every family $f$ has equal weight regardless of its number of cells $N_f$; the family label is used only for this weighting and is never a model input.
For gapped cells the two spectra are aligned at their VBM, and $\ell_{\rm win}$ is the MAE over reference states within $\pm2$\,eV of the midgap, with the VBM and CBM bands always included and weighted by 5.
The gap term $\ell_{\rm gap}=|E_g^{\rm EP}-\max(E_g^{\rm PBE},0)|$ uses the \emph{signed} predicted gap: an unclamped gap keeps a gradient when the model closes it, which a clamped gap does not.
For metals the spectra are aligned at a Fermi-level proxy, the midpoint between the highest occupied and lowest unoccupied band, and $\ell_{\rm win}$ is the MAE within $\pm2$\,eV of it.
The deep-valence term $\ell_{\rm deep}$ is the MAE, with the same alignment, of the occupied states whose reference energy lies below $E_{\rm ref}-2$\,eV, where $E_{\rm ref}$ is the midgap (gapped cells) or the Fermi-level proxy (metals) at the centre of the window; these are exactly the occupied states below the window's lower edge. The term corrects the occupied bandwidth, which the window alone leaves unconstrained.
We use $\lambda_g=1$ and $\lambda_d=0.1$.

Training uses AdamW~\cite{Loshchilov2019} with full batches, and the cells fall into 15 loss groups.
Three parameter groups get their own learning rates: the environment networks (encoder, readouts, router), the species potential, and the $s/p$ strengths with their head.
Weight decay is $10^{-4}$ for the first two groups and zero for the third.
The 66-cell precursor was trained from scratch for 1700 steps in three stages of 1000, 400 and 300 steps; EpNet-Si was then trained on all 77 cells starting from the precursor, with the last two stages only (700 steps; training history in Fig.~\ref{fig:history} of Appendix~\ref{app:history}).
The environment networks use learning rates of $3\times10^{-3}$, $10^{-3}$ and $3\times10^{-4}$ in the three stages, the species potential $10^{-3}$, $3\times10^{-4}$ and $10^{-4}$, and the $s/p$ group $10^{-2}$, $3\times10^{-3}$ and $10^{-3}$.
The optimizer is reinitialized at each stage.
Every second $k$ point of each path is used in the loss.
A step takes 28--35\,s on one NVIDIA L40S GPU, about 13\,h for the precursor and 7\,h for the final stages; diagonalizing the largest cage cells (about 3150 plane waves) accounts for most of the step time.

\subsection{Evaluation protocol}
\label{sec:protocol}

A gapped cell (path gap $>0.02$\,eV) \emph{passes} if its band MAE is $\le100$\,meV and its gap error $\le50$\,meV.
It is a \emph{near miss} if the band MAE is $\le100$\,meV and $50<|\Delta E_g|\le100$\,meV, and it \emph{fails} otherwise.
A metal is scored on the uniform grid, with the PBE $E_F$ fixed by electron count and the model shifted by its median offset within $\pm1$\,eV.
It passes if the grid band MAE is $\le50$\,meV, the relative DOS difference $\int|g-g_{\rm PBE}|\,dE/\int g_{\rm PBE}\,dE$ within $\pm1$\,eV of $E_F$ (Gaussian broadening 0.1\,eV) is $\le5\%$, and at most 0.5\% of electrons are misplaced.
It fails if the band MAE exceeds 100\,meV or the DOS difference 10\%, and is a near miss otherwise.


\subsection{Derived properties}
\label{sec:derived-methods}
The loss constrains band energies on the reference $k$ paths, but most uses of a band structure rely on quantities derived from it.
We therefore test four that the loss never sees: the curvature of the bands at their edges (effective masses), their response to volume (pressure coefficients of the gap) and to anisotropic strain (deformation potentials), and, for metals, the density of states at the Fermi level $N(E_F)$.

\emph{Effective masses.} The label paths are too coarse to resolve band curvature, so new QE band calculations along straight lines through the PBE band extrema, $|t|\le0.12$\,\AA$^{-1}$ in steps of 0.004\,\AA$^{-1}$ ($|t|\le0.07$\,\AA$^{-1}$ for cage cells), reuse the converged SCF density; the model is evaluated on the same lines.
Each line passes through the VBM or CBM found on the label path, $\mathbf k_0+t\hat{\mathbf u}$, with $\hat{\mathbf u}$ along each of the path segments that leave that point (one or two distinct directions per edge), so every mass is a directional mass along a high-symmetry direction rather than a full mass tensor.
We fit $E=E_0+a(t-t_0)^2$ within 0.03\,\AA$^{-1}$ of each line's extremum and take $|m^*|/m_e=\hbar^2/(2m_e|a|)$.
Electrons are fitted on the lowest conduction band and holes on the top three valence bands; for the second and third valence bands, which need not be degenerate with the VBM, we keep only lines on which both the PBE and the model extremum lie within $10^{-3}$\,\AA$^{-1}$ of $\mathbf k_0$, and report them as valence-band curvatures at the VBM wavevector.

\begin{figure*}[t]
\includegraphics[width=\textwidth]{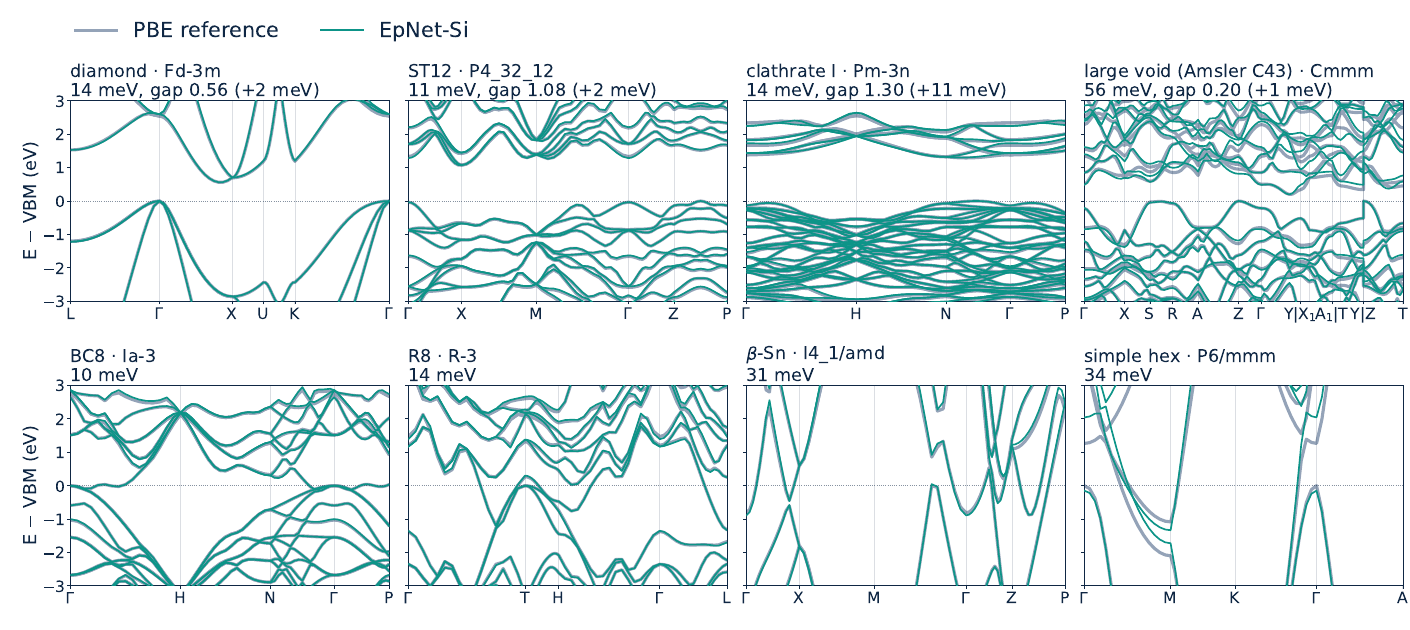}
\caption{\label{fig:train}Eight training cells: QE-PBE (grey) and EpNet-Si (teal). Top, semiconductors (diamond, ST12, clathrate~I and a large-void Amsler framework); bottom, metals (BC8, R8, $\beta$-Sn, simple hexagonal). Gapped cells are aligned at the VBM and metals at $E_F$; titles give the band error and the PBE gap with its error.}
\end{figure*}

\emph{Pressure coefficients.} The response of the gap to hydrostatic compression is obtained from seven cells per family between $-6$ and $+6\%$ volume (internal coordinates scaled, not relaxed), of which at most three (the parent and the $\pm4\%$ cells) are in training.
A linear fit of the path gap in $\ln V$ gives the volume deformation potential $a_g=dE_g/d\ln V$.
The bulk modulus $B_0$ comes from a third-order Birch--Murnaghan fit~\cite{Birch1947} to the PBE total energies, and $dE_g/dP=-a_g/B_0$, with the PBE $B_0$ used for both the model and the reference.

\emph{$N(E_F)$.} For metals, the density of states at the Fermi level controls screening and transport.
On the uniform grid, $E_F$ is found by electron count with 0.05\,eV Gaussian smearing, in the same way for model and reference, and the DOS is broadened by 0.1\,eV.

\emph{Deformation potentials.} 
For 40 selected cells under uniaxial and shear strain, QE eigenvalues were computed at the band-edge $k$ star of the parent, found with spglib~\cite{spglib}, and the strain was obtained from the polar decomposition of the deformation gradient $F=RU$, where $F$ maps the parent lattice vectors onto the strained ones, $R$ is a rotation and $U=(F^{\mathsf T}F)^{1/2}$ is the symmetric stretch; the strain tensor is the Biot strain $\varepsilon=U-I$.
Absolute band-edge positions are not comparable between cells, so we compare the strain-induced splittings of the conduction valleys and of the valence-band maximum, and the change of the gap.
For diamond the splittings are fitted with Bir--Pikus expressions to obtain $\Xi_u$ and $b$~\cite{Fischetti1996}.

\section{Results}
\label{sec:results}

In this section, we evaluate EpNet-Si on 77 training cells (Sec.~\ref{sec:res-train}) and 423 structures outside training (Sec.~\ref{sec:res-groups}).
In brief, distortions of the trained structure types are reproduced at training accuracy with no failures, whereas structure types the model has never seen keep their PBE band shapes but carry conduction-band offsets, largest in the most open frameworks.
We then ask which training data matter by leaving out one family at a time (Sec.~\ref{sec:res-lofo}), and test quantities derived from the bands (Sec.~\ref{sec:res-derived}); what the trained network has learned is discussed in Sec.~\ref{sec:res-interp}.


\subsection{Training set}
\label{sec:res-train}

Fig.~\ref{fig:train} shows the fit on eight training cells.
Over the 77 cells the band error is 23\,meV and the gap error 5\,meV, and all 48 training gaps lie within 50\,meV of PBE (largest 26\,meV).
The fourfold semiconductors are fitted at 10--20\,meV and the clathrate~I cages at 17\,meV with a 7\,meV gap error.
The largest band errors belong to the three large-void cells (55\,meV) and to the one- and two-atom metals, $\beta$-Sn and simple hexagonal (32--40\,meV).

\subsection{Test groups}
\label{sec:res-groups}

\begin{figure}[t]
\includegraphics[width=0.9\columnwidth]{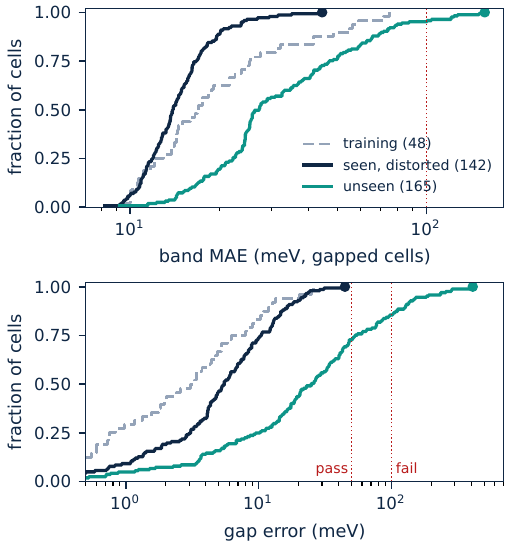}
\caption{\label{fig:groups}Cumulative distributions of the band MAE (top) and gap error (bottom) over the gapped cells of the two test groups, with the 48 gapped cells of the 77 training cells for reference (dashed). Dotted lines mark the pass (50\,meV) and fail (100\,meV) gap thresholds and the 100\,meV band threshold; dots mark each group's worst cell.}
\end{figure}

Table~\ref{tab:groups} and Fig.~\ref{fig:groups} give the errors on all 423 test structures.
On distortions of the trained structure types the model is as accurate as on its training set: 15\,meV band and 8\,meV gap error; every gapped cell passes, and 88 of the 89 metals pass with one near miss.
On structure types it has never seen the errors are larger, 38\,meV band and 49\,meV gap error with a median gap error of 25\,meV; 73\% of the gapped cells pass and 15\% fail.
Metals of both groups are reproduced at 15--21\,meV near $E_F$.

\begin{figure*}[t]
\includegraphics[width=\textwidth]{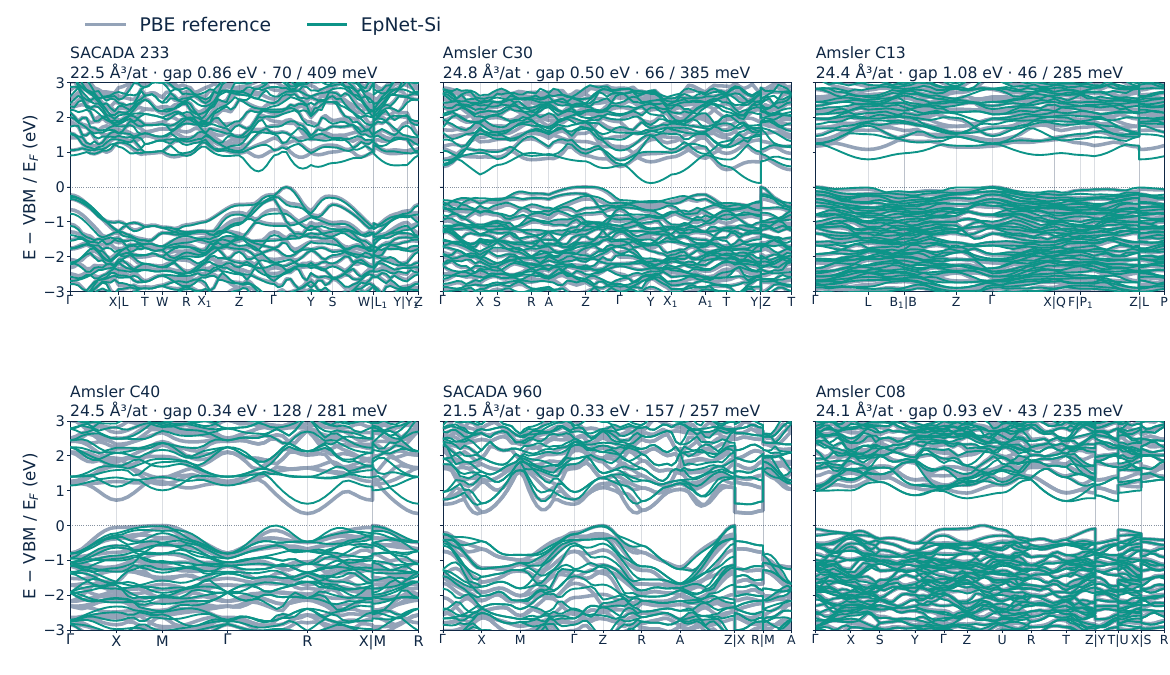}
\caption{\label{fig:worst}The six unseen structures with the largest gap errors: PBE (grey) and EpNet-Si (teal), VBM-aligned. Titles give the volume per atom, the PBE gap and the band / gap errors (meV).}
\end{figure*}

\begin{table}[!ht]
\caption{\label{tab:groups}EpNet-Si on the 423 test structures. Grades as in Sec.~\ref{sec:protocol}; the worst cell is the largest gap error of each group.}
\begin{ruledtabular}
\begin{tabular}{lrr}
 & Seen, distorted & Unseen \\
\hline
Structures & 231 & 192 \\
Gapped cells & 142 & 165 \\
\quad Band / gap (meV) & 15 / 8 & 38 / 49 \\
\quad Median gap error (meV) & 6 & 25 \\
\quad Pass / near miss / fail & 142 / 0 / 0 & 120 / 20 / 25 \\
Metals & 89 & 27 \\
\quad Band (meV) & 15 & 21 \\
\quad Pass / near miss / fail & 88 / 1 / 0 & 22 / 5 / 0 \\
Worst gap error (meV) & 45 & 409 \\
\end{tabular}
\end{ruledtabular}
\end{table}

\emph{Distortions of seen structures.}
Fig.~\ref{fig:known} in Appendix~\ref{app:seen} shows examples from eight trained structure types, selected by name rather than score.
All 95 gapped single distortions (15 / 8\,meV) and 47 combined distortions (15 / 7\,meV) pass with similar errors, indicating that combining strain, shear and rattling does not reduce accuracy.
The largest gap error is 45\,meV for rattled lonsdaleite.

\emph{Unseen structures.}
New polytypes, clathrate~II and OQMD structures transfer near training accuracy (16--26\,meV band error); all 22 clathrate~II cells pass despite no cage of this type being trained.
SACADA, Amsler and AFLOW frameworks account for all 25 failures and 19 of 20 near misses (42--45\,meV band, 54--76\,meV gap error).
Their dispersions follow PBE (Fig.~\ref{fig:new}, Appendix~\ref{app:unseen}), with conduction-manifold offsets mostly of 0.05--0.25\,eV.
The six worst gap errors (Fig.~\ref{fig:worst}) are 0.24--0.41\,eV, dominated by the lowest conduction band: four gaps are underestimated and two (Amsler C40 and SACADA 960) overestimated.
Four are Amsler frameworks at 24.1--24.8\,\AA$^3$/atom, at or beyond the most open training framework (24.0\,\AA$^3$/atom).
The worst, SACADA net 233, has a larger void than clathrate~I (3.48 versus 3.25\,\AA) and a predicted gap of 0.45 versus 0.86\,eV in PBE. Its conduction-band minimum has a density overlap of 0.97 with PBE, but 17\% of its weight lies in the void region versus 6\% in PBE. Here the densities $\rho(\mathbf r)=|\psi(\mathbf r)|^2$ of the lowest conduction band of both methods, evaluated at the same wavevector (the PBE CBM on the label path; PBE pseudo-wavefunction), are sampled on the same $26^3$ grid over the cell and normalized to $\sum_{\mathbf r}\rho=1$; the overlap is $S=\sum_{\mathbf r}\sqrt{\rho_{\mathrm{PBE}}\,\rho_{\mathrm{EP}}}$, equal to 1 for identical densities, and the void weight is $\sum\rho$ over the grid points farther than 2.25\,\AA\ from every atom, including periodic images.
The six worst seen-structure distortions are shown in Fig.~\ref{fig:knownworst}, Appendix~\ref{app:overlays}.

\subsection{Transferability across bonding environments}
\label{sec:res-lofo}

\begin{figure}[t]
\includegraphics[width=0.85\columnwidth]{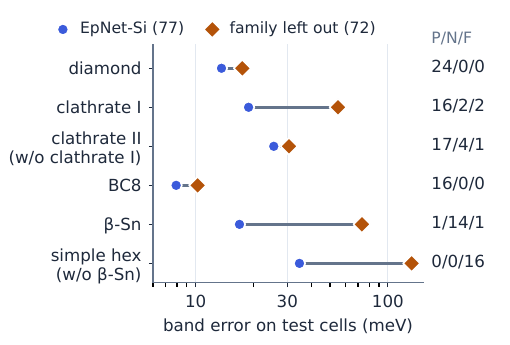}
\caption{\label{fig:lofo}Transferability across bonding environments. Test band errors for EpNet-Si (circles) and models retrained without the indicated family (diamonds). ``w/o'' identifies the omitted family for related structures. Right: pass/near miss/fail counts for the retrained models.}
\vspace{-3mm}
\end{figure}

To find which training data matter most, we retrained the recipe four times, each time without all cells of one family: diamond, clathrate~I, BC8 or $\beta$-Sn (Fig.~\ref{fig:lofo}).
Each run repeats the two-step training of EpNet-Si without that family: from scratch on the 66-cell precursor set, then on the 77-cell set, so it never sees any cell of the family.
Each run was then scored on that family's test distortions.
Diamond and BC8 can be left out: all 40 of their distortions still pass, with band errors rising only from 14 to 17 and from 8 to 10\,meV, because families with the same local bonding remain in training (4H, lonsdaleite and oC12; R8 and tI16).
Clathrate~I cannot.
Without it, the band error on its distortions rises from 19 to 55\,meV and the gap error from 8 to 42\,meV, with 2 of 20 cells failing and 2 near misses. For the never-trained clathrate~II, the band error rises from 26 to 31\,meV and the gap error from 15 to 33\,meV, with 1 failure and 4 near misses.
Nor can $\beta$-Sn: without it, the band error on its metallic distortions rises from 17 to 73\,meV (14 of 16 cells are near misses and 1 fails). For simple hexagonal cells, which remain in training, the band error rises from 35 to 133\,meV and all 16 fail.
As a control, band errors on all other test cells remain nearly unchanged across the four runs: 15\,meV for seen structures and 36--38\,meV for unseen structures.
These tests show that some families with similar local bonding provide redundant training information.
Clathrate~I and $\beta$-Sn provide distinct bonding environments that the remaining training families cannot replace.
Only four of the twelve families were tested in this way.

\subsection{Derived properties}
\label{sec:res-derived}

\begin{figure}[t]
\includegraphics[width=\columnwidth]{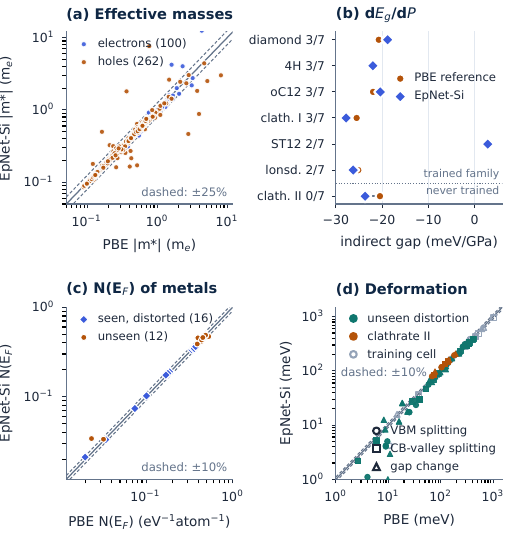}
\caption{\label{fig:derived}Properties outside the training loss. (a) 362 effective masses from 7 parents and 66 gapped development cells; dashed lines $\pm25\%$. (b) Gap pressure coefficients, $dE_g/dP=-a_g/B_0$; labels count training cells among 7 hydrostatic cells per family. (c) $N(E_F)$ for 28 metals outside training. (d) Magnitudes of valence-band splittings (circles), conduction-valley splittings (squares) and gap changes (triangles) for 40 strained cells of seven structures. Dashed lines in (c,d): $\pm10\%$. Gap-change signs match PBE except one ST12 cell ($+1$ versus $-10$\,meV).}
\end{figure}

We compared four derived properties, none used directly as training targets, with PBE: effective masses, the pressure coefficient of the gap, $N(E_F)$ of metals, and uniaxial and shear deformation potentials (Fig.~\ref{fig:derived}).

\emph{Effective masses.} We computed PBE and EpNet-Si bands through the band edges of the 7 gapped family parents and of 66 gapped development-set cells, and fitted 362 directional masses, 100 for electrons and 262 for holes, from 73 cells (Fig.~\ref{fig:derived}a).
The median error is 3.1\% and 93\% of the masses are within 25\%.
For diamond, EpNet-Si gives a longitudinal electron mass $m_l=0.950\,m_e$ against 0.956 in PBE.
A notable discrepancy occurs at the conduction-band minimum of clathrate~II, where the predicted band is too flat and the effective mass is substantially overestimated.

\begin{figure*}[t]
\includegraphics[width=0.8\textwidth]{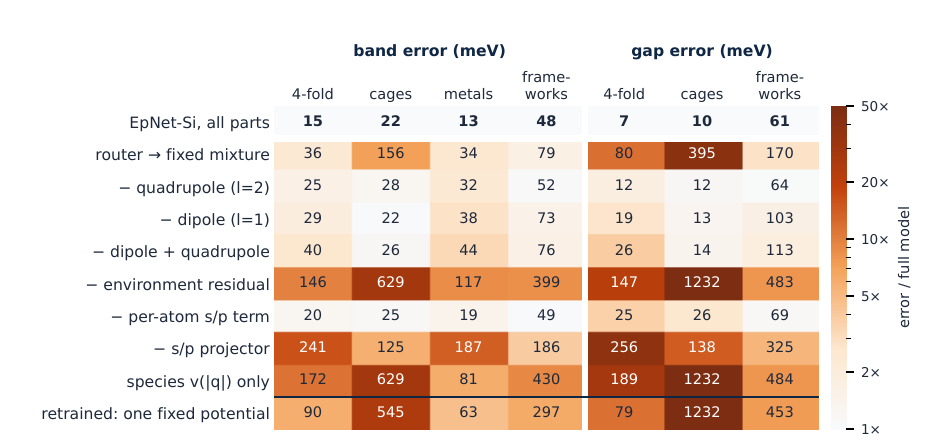}
\caption{\label{fig:ablation}Component ablation on the 84 structures of the development set (both test groups). Each row switches off one part of the trained EpNet-Si at inference, without retraining. The last row, below the rule, is a baseline with one fixed potential for every atom (species $v(|q|)$ and $s/p$ strengths only), retrained on the same 77 cells and scored the same way. Entries are mean band and gap errors (meV) per structure class; color gives the ratio to the full model. ``Router $\to$ fixed mixture'' replaces $\pi_{a,e}$ by its training mean for every atom; ``species $v(|q|)$ only'' is the classical EPM limit.}
\end{figure*}

\emph{Pressure coefficient of the gap.} From seven hydrostatic cells per family, at most three of them in training, we obtained $a_g=dE_g/d\ln V$ and converted it to $dE_g/dP=-a_g/B_0$ with the PBE bulk modulus of the same cells (89\,GPa for diamond, 78\,GPa for the clathrates).
Fig.~\ref{fig:derived}b compares the two for seven families.
EpNet-Si reproduces the positive sign of ST12's coefficient ($+2.8$ against $+3.0$\,meV/GPa), opposite to the other fourfold phases.
4H and lonsdaleite are within 2--4\% of PBE and ST12 within 6\%, diamond and oC12 are 7--9\% too shallow, clathrate~I is 9\% too steep, and the never-trained clathrate~II is 16\% too steep ($-23.7$ against $-20.5$\,meV/GPa).
The compressed cell in each family matters here: an earlier 66-cell model, trained with only the expanded hydrostatic cell, was 30\% too steep for clathrate~I and 45\% for clathrate~II, because compression was an extrapolation.

\emph{Density of states at $E_F$.} For the 28 metals outside training, $N(E_F)$ has a median relative error of 1.7\% against PBE, and 25 of the 28 cells are within 10\% (Fig.~\ref{fig:derived}c). The 16 distortions of seen structures all have relative errors below 8\% (median 2.0\%); the 12 unseen structures have a median relative error of 1.5\%. The three cells with errors above 10\% are a SACADA framework metal (46\%) and two OQMD metals (11 and 15\%).

\emph{Deformation potentials.} Unlike hydrostatic pressure, uniaxial and shear strains lower crystal symmetry and split degenerate band edges.
In diamond Si, these splittings affect the conduction valleys near $X$ and the heavy- and light-hole bands at $\Gamma$, underpinning the mobility gain in strained-silicon transistors~\cite{Sun2007}; deformation potentials quantify their strain response.
We strained seven fourfold and cage structures (diamond, 4H, lonsdaleite, oC12, ST12, clathrate~I and clathrate~II; 40 cells) and compared, at the band-edge $k$ star of each parent, three quantities: the splitting of the valence-band maximum, the splitting of the conduction valleys and the change of the gap (Fig.~\ref{fig:derived}d).
The PBE medians are 65, 34 and 95\,meV; for the 30 strained test cells of trained families, EpNet-Si has median relative errors of 6\%, 8\% and 5\%, respectively.
Splittings below 30\,meV in ST12 and clathrate~I show greater relative scatter, but absolute errors remain within 11\,meV.
Errors rise to 10--13\% for the four cells of unseen clathrate~II.
For diamond, Bir--Pikus fits give $\Xi_u=7.9$\,eV and $b=-2.59$\,eV, compared with PBE values of 8.1 and $-2.40$\,eV, differing by about 2.5\% and 8\%, respectively.

\section{Discussion}
\label{sec:discussion}

\subsection{What the network learns}
\label{sec:res-interp}

\begin{figure*}
\includegraphics[width=0.9\textwidth]{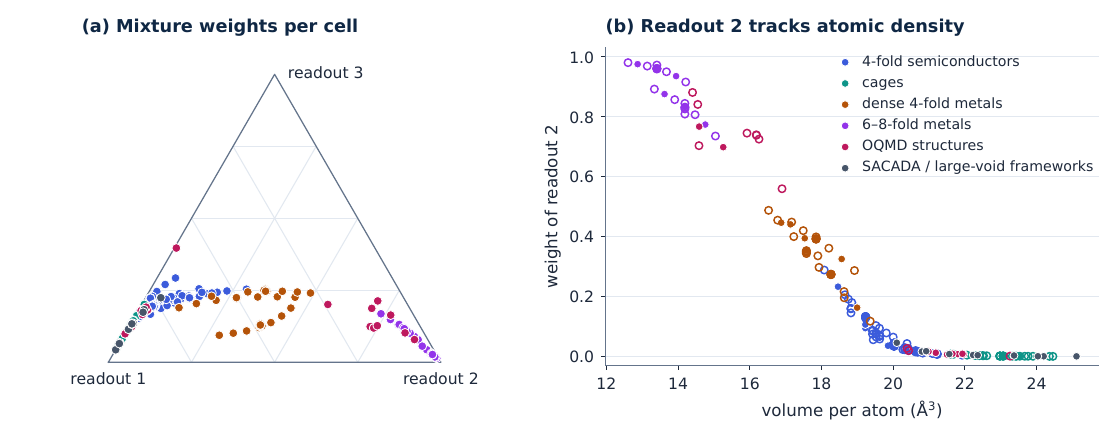}
\caption{\label{fig:router}Mixture weights $\pi_{a,e}$ of the density router, averaged over the atoms of each cell, for 77 training cells and 173 test cells (open markers). (a) Ternary plot of the three readout weights. (b) Weight of readout 2 versus volume per atom.}
\end{figure*}

\emph{Which components matter.} To see what the trained model relies on, we switched off one component at a time at inference and rescored the 84 structures of the development set (Fig.~\ref{fig:ablation}).
Because the components were trained together, this measures their contribution within the co-adapted model; the last row, one fixed potential retrained on the same 77 cells (training loss 0.62, against 0.03 for EpNet-Si), shows that a model without them fails as well.
The environment residual and the shared $s/p$ projector are both essential: without either, the errors grow by factors of roughly 4--120, and the species potential alone misses the cage gaps by 1.2\,eV and the framework gaps by 0.48\,eV.
Replacing the router by a fixed, atom-independent mixture raises the fourfold gap error from 7 to 80\,meV and the cage gap error from 10 to 395\,meV; mixing the readouts per atom is what lets one encoder describe both open and dense silicon.
The dipole and quadrupole terms raise the band and gap errors of fourfold and metallic cells by factors of 1.6--4 when removed, and the framework gap error from 61 to 64--113\,meV, but the cage gaps barely need them (10 to 12--14\,meV).
The per-atom correction to the $s/p$ strengths is a smaller refinement: without it the gap errors of the fourfold phases and cages rise from 7 to 25 and from 10 to 26\,meV.

\begin{figure*}[t]
\includegraphics[width=0.9\textwidth]
{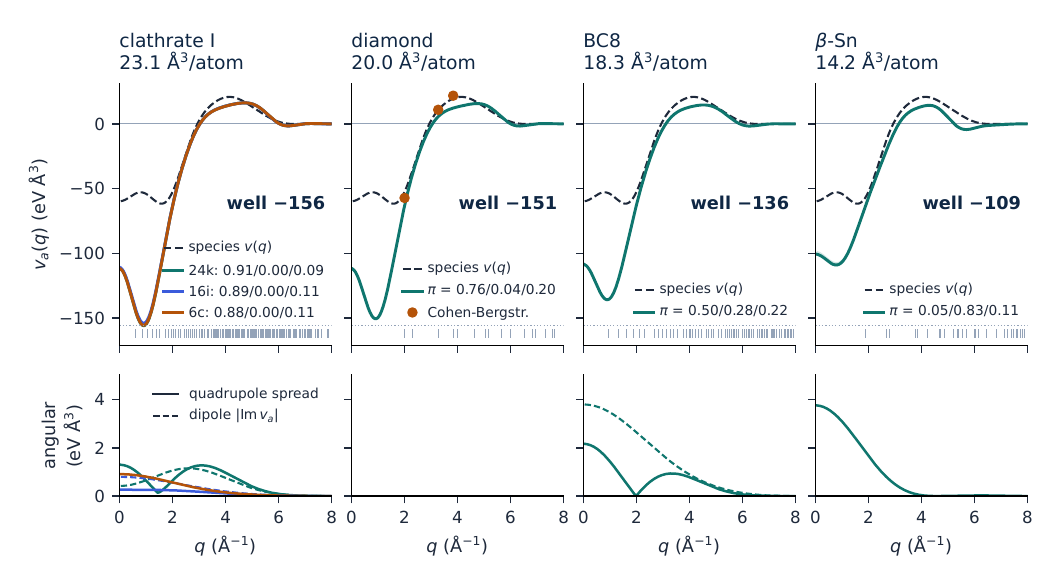}
\vspace{-4mm}
\caption{\label{fig:formfactor}Learned form factors $v_a(\mathbf G)$, Eq.~\eqref{eq:va}, on every symmetry-distinct site of four parents, ordered by decreasing volume per atom. Top: isotropic part (legend: Wyckoff site and router weights $\pi_{a,1}/\pi_{a,2}/\pi_{a,3}$), species potential $v(q)$ (dashed), well depth near 1\,\AA$^{-1}$ (dotted line: clathrate~I value), sampled $|\mathbf G|$ shells (ticks) and Cohen--Bergstresser values (dots, diamond). Bottom: angular part over 400 directions of $\mathbf G$: quadrupole spread of ${\rm Re}\,v_a$ (solid) and largest dipole $|{\rm Im}\,v_a|$ (dashed).}
\end{figure*}

\begin{table*}[t]
\caption{\label{tab:design}Development path from the Cohen--Bergstresser EPM to EpNet-Si.}
\begin{ruledtabular}
\begin{tabular}{p{0.25\textwidth}p{0.34\textwidth}p{0.34\textwidth}}
Step & What it adds & What it fixed, and what broke \\
\hline
Cohen--Bergstresser $V_S$ & three form factors on fixed shells & diamond only \\
Smooth species $v(|q|)$ & 12 cosine coefficients, $q_{\rm cut}=8$\,\AA$^{-1}$, any cell & clathrate~I becomes a metal \\
+ equivariant $l\le2$ residual & per-atom monopole, dipole and quadrupole from the local environment & isotropic cages still wrong \\
+ $s/p$ nonlocal projector & four Kleinman--Bylander strengths, giving a $\mathbf k$ dependence no local $V$ can & every family improves by 0.03--0.05\,eV\\
+ hand-set packing gates & coordination-number windows route cage atoms to a separate branch & cage gaps open\\
+ continuous density router & one encoder, three readouts mixed by a learned softmax of local densities; signed gap loss & fresh distortions 0.017 / 0.007\,eV (band / gap); first external test: framework gaps miss \\
+ $k$ folding, deep-valence loss & $\mathbf k$ folded into the first zone; loss on occupied bands below the window & deep-valence error $0.166\to0.034$\,eV\\
+ modified training set (EpNet-Si) & final set (Table~\ref{tab:data}): 56 family cells (up to five each), 11 OQMD structures, seven other frameworks, three large-void cells; 143 $\to$ 77 cells & distortions unchanged; unseen-framework gap error $0.087\to0.061$\,eV; clathrate~I pressure slope 30\% $\to$ 9\% off \\
+ $k$-centred basis, smooth buffer & $|\mathbf k+\mathbf G|\le G_{\rm max}$ with a decoupling buffer to $1.05\,G_{\rm max}$; readouts vanish at $q_c$ & same bands in any supercell (diamond gap: 32\,meV cell dependence $\to$ 0); $E(\mathbf k)$ continuous \\
\end{tabular}
\end{ruledtabular}
\end{table*}

\emph{The role of the router.} According to Eq.~\eqref{eq:readout}, each atom receives three candidate residual form factors, mixed by a router that sees only local atomic densities. No roles were assigned to the readouts, yet their meaning can be read off from where the router sends each environment (Fig.~\ref{fig:router}a).
Readout 1 serves open covalent environments (weight 0.90 in cages, about 0.76 in fourfold semiconductors), readout 2 prefers dense metallic ones (0.83--0.96 in the sixfold to eightfold metals, nearly zero above 22\,\AA$^3$/atom), and readout 3 acts as a covalent correction (about 0.2 in fourfold phases, 0.1 in cages and metals).
The dense fourfold metals BC8, R8 and tI16 mix all three, as befits their intermediate bonding.
More interestingly, Fig.~\ref{fig:router}b shows that the weight of readout 2 follows a single smooth curve in volume per atom, and the 173 test cells lie on the same curve as the training cells. These test cells are all the non-training cells stored with the training data, for which router weights were recorded during training: 77 left-out catalog distortions, 85 cells of earlier reference sets (distortions of trained families and 15 clathrate~II cells) and 11 held-out OQMD structures; they were not selected by error. The router has thus recovered an open/compact/metallic classification from local densities alone. 

\emph{The learned form factors.} Fig.~\ref{fig:formfactor} shows $v_a(q)$ on every symmetry-distinct site of four parents.
The largest departure from the species potential is a deep attractive well near 1\,\AA$^{-1}$.
Ordered by volume per atom, the well becomes steadily shallower as the structure becomes denser: $-156$\,eV\,\AA$^3$ in clathrate~I (23.1\,\AA$^3$/atom), $-151$ in diamond (20.0), $-136$ in BC8 (18.3) and $-109$ in $\beta$-Sn (14.2), continuing to $-105$ in simple hexagonal Si (13.4). No global volume-per-atom or structure-type label enters the readouts; the trend arises because the router weights vary smoothly with the local density (Fig.~\ref{fig:router}b). Its direction agrees with the long-wavelength limit of a screened pseudopotential, $v(q\to0)=-\tfrac23E_F\,\Omega_{\rm at}$, which for a free-electron gas of four electrons per atom deepens with the volume per atom as $\Omega_{\rm at}^{1/3}$~\cite{Harrison1966}, although the learned trend is stronger and the region below the shortest sampled $|\mathbf G|$ is not constrained by the data.
The angular part (bottom row of Fig.~\ref{fig:formfactor}) is small, at most 3.8\,eV\,\AA$^3$ against isotropic changes up to 100\,eV\,\AA$^3$. It vanishes on the $T_d$ site of diamond. The dipole also vanishes on the $\beta$-Sn and clathrate-6c sites, whose $D_{2d}$ symmetry admits no vector, and is largest on the polar threefold site of BC8; the quadrupole is largest on the $\beta$-Sn site, with its strongly anisotropic 4+2 coordination.

Together, these analyses relate the model's learned environment dependence to identifiable corrections to the empirical pseudopotential Hamiltonian.

\subsection{What makes EpNet-Si transferable}
\label{sec:design}

Table~\ref{tab:design} traces the development of EpNet-Si through iterative testing and validation to its final form. Importantly, two lessons stand out.

First, a smooth, spherical species potential already transfers across eight families at about 0.1\,eV, but it cannot open the gap of the clathrate cages; what it lacks is not angular resolution but a $\mathbf k$-dependent, orbital-selective term.
The equivariant dipole and quadrupole channels vanish on sites with $T_d$ symmetry, such as those of diamond, and the decisive improvement came from the $s/p$ projector, which acts differently on $s$-like and $p$-like plane-wave combinations and so adjusts the $s$--$p$ splitting that sets the gaps of $sp^3$ networks.
The ablation of the trained model supports this reading: switching off the shared $s/p$ strengths raises the errors of every structure class by factors of 4--37, more than removing both angular channels (Fig.~\ref{fig:ablation}).

Second, routing design is critical. We first used a discrete router: coordination-number gates sent cage atoms to their own branch, which opened the cage gaps, but failures on an external evaluation set revealed sensitivity to atoms assigned to an inappropriate branch.
Replacing the gates by a continuous router over local densities removed that failure mode, and the learned router reconstructs the density axis on its own (Fig.~\ref{fig:router}).
The router is not cosmetic: fixing the mixture weights to their training mean at inference raises the cage gap error almost 40-fold.

\subsection{Limitations and extensions}

EpNet-Si is trained on PBE band structures and so inherits the PBE band-gap error. For diamond Si the target gap is 0.56\,eV, not the experimental 1.17\,eV~\cite{Perdew1983}.
Since the model is trained on band energies alone, it can be retrained on $GW$~\cite{Hybertsen1986} or experimental targets without architectural change, as DeepPseudopot does for zinc-blende semiconductors~\cite{Lin2025}.
Spin--orbit coupling, which is small in Si, can enter through the existing $s/p$ projectors.
The largest remaining errors occur in unseen frameworks: four of the six worst cases have volumes per atom at or beyond the most open training framework, while the worst case, SACADA net 233, has a void larger than clathrate~I's but within the range of the large-void training cells.
Their band shapes are reproduced, but conduction-band offsets can either underestimate or overestimate the gap, with errors up to 0.41\,eV.
One possible explanation is that fitting band energies alone leaves the potential in large voids insufficiently constrained; the present results do not establish this mechanism or separate it from limited coverage of open environments in training.
Testing additional open-framework training cells and constraints on the potential inside voids could help distinguish these explanations.
Compounds require several species potentials and a treatment of charge transfer, which the present single-species model does not address; defects and surfaces require supercells large enough that iterative eigensolvers~\cite{Wang1994} replace dense diagonalization.
The differentiable Hamiltonian offers routes to band velocities, effective masses and deformation potentials, and potentially to optical matrix elements; it may also provide a starting point for future studies of electron--phonon coupling.

\section{Conclusions}
\label{sec:conclusion}

In this work, we have shown that an empirical pseudopotential becomes transferable across bonding environments once its form factors depend continuously on each atom's surroundings.
EpNet-Si realizes this with an equivariant network that predicts per-atom plane-wave form factors, a learned density router in place of hand-set thresholds, and a small $s/p$ nonlocal term.
Trained on only 77 cells with diverse local environments, one model is tested on 231 distortions of seen structures and 192 unseen structures.
Mean band and gap errors remain within a few tens of meV in both groups, with larger gap errors concentrated in unseen open frameworks.
The accuracy carries over to properties derived from the bands: effective masses, deformation potentials, the gap pressure coefficients of the trained families, and $N(E_F)$ of metals outside training.
Switching components off at inference shows that the trained model relies on the residual form factor, the $s/p$ projector and the per-atom density router, and the router itself recovers an open/compact/metallic classification from local densities.
These results suggest that the EPM's historical weakness, transferability, can be addressed by learning the environment dependence of the potential from a small, diverse set of band structures.
\begin{acknowledgments}
This research was sponsored by the U.S. Department of Energy, Office of Science, Office of Basic Energy Sciences, and the Established Program to Stimulate Competitive Research (EPSCoR) under the DOE Early Career Award No. DE-SC0024866. The computing resources are provided by ACCESS (TG-MAT230046). 
\end{acknowledgments}

\section*{Data availability}
The EpNet-Si source code, the trained model, and the scripts used to produce the results of this study are available at \url{https://github.com/MaterSim/EpNet-Si}.

\bibliography{refs}

\begin{thebibliography}{49}%
\makeatletter
\providecommand \@ifxundefined [1]{%
 \@ifx{#1\undefined}
}%
\providecommand \@ifnum [1]{%
 \ifnum #1\expandafter \@firstoftwo
 \else \expandafter \@secondoftwo
 \fi
}%
\providecommand \@ifx [1]{%
 \ifx #1\expandafter \@firstoftwo
 \else \expandafter \@secondoftwo
 \fi
}%
\providecommand \natexlab [1]{#1}%
\providecommand \enquote  [1]{``#1''}%
\providecommand \bibnamefont  [1]{#1}%
\providecommand \bibfnamefont [1]{#1}%
\providecommand \citenamefont [1]{#1}%
\providecommand \href@noop [0]{\@secondoftwo}%
\providecommand \href [0]{\begingroup \@sanitize@url \@href}%
\providecommand \@href[1]{\@@startlink{#1}\@@href}%
\providecommand \@@href[1]{\endgroup#1\@@endlink}%
\providecommand \@sanitize@url [0]{\catcode `\\12\catcode `\$12\catcode
  `\&12\catcode `\#12\catcode `\^12\catcode `\_12\catcode `\%12\relax}%
\providecommand \@@startlink[1]{}%
\providecommand \@@endlink[0]{}%
\providecommand \url  [0]{\begingroup\@sanitize@url \@url }%
\providecommand \@url [1]{\endgroup\@href {#1}{\urlprefix }}%
\providecommand \urlprefix  [0]{URL }%
\providecommand \Eprint [0]{\href }%
\providecommand \doibase [0]{https://doi.org/}%
\providecommand \selectlanguage [0]{\@gobble}%
\providecommand \bibinfo  [0]{\@secondoftwo}%
\providecommand \bibfield  [0]{\@secondoftwo}%
\providecommand \translation [1]{[#1]}%
\providecommand \BibitemOpen [0]{}%
\providecommand \bibitemStop [0]{}%
\providecommand \bibitemNoStop [0]{.\EOS\space}%
\providecommand \EOS [0]{\spacefactor3000\relax}%
\providecommand \BibitemShut  [1]{\csname bibitem#1\endcsname}%
\let\auto@bib@innerbib\@empty
\bibitem [{\citenamefont {Phillips}\ and\ \citenamefont
  {Kleinman}(1959)}]{Phillips1959}%
  \BibitemOpen
  \bibfield  {author} {\bibinfo {author} {\bibfnamefont {J.~C.}\ \bibnamefont
  {Phillips}}\ and\ \bibinfo {author} {\bibfnamefont {L.}~\bibnamefont
  {Kleinman}},\ }\bibfield  {title} {\bibinfo {title} {New method for
  calculating wave functions in crystals and molecules},\ }\href
  {https://doi.org/10.1103/PhysRev.116.287} {\bibfield  {journal} {\bibinfo
  {journal} {Phys. Rev.}\ }\textbf {\bibinfo {volume} {116}},\ \bibinfo {pages}
  {287} (\bibinfo {year} {1959})}\BibitemShut {NoStop}%
\bibitem [{\citenamefont {Cohen}\ and\ \citenamefont
  {Heine}(1961)}]{Cohen1961}%
  \BibitemOpen
  \bibfield  {author} {\bibinfo {author} {\bibfnamefont {M.~H.}\ \bibnamefont
  {Cohen}}\ and\ \bibinfo {author} {\bibfnamefont {V.}~\bibnamefont {Heine}},\
  }\bibfield  {title} {\bibinfo {title} {Cancellation of kinetic and potential
  energy in atoms, molecules, and solids},\ }\href
  {https://doi.org/10.1103/PhysRev.122.1821} {\bibfield  {journal} {\bibinfo
  {journal} {Phys. Rev.}\ }\textbf {\bibinfo {volume} {122}},\ \bibinfo {pages}
  {1821} (\bibinfo {year} {1961})}\BibitemShut {NoStop}%
\bibitem [{\citenamefont {Cohen}\ and\ \citenamefont
  {Bergstresser}(1966)}]{Cohen1966}%
  \BibitemOpen
  \bibfield  {author} {\bibinfo {author} {\bibfnamefont {M.~L.}\ \bibnamefont
  {Cohen}}\ and\ \bibinfo {author} {\bibfnamefont {T.~K.}\ \bibnamefont
  {Bergstresser}},\ }\bibfield  {title} {\bibinfo {title} {Band structures and
  pseudopotential form factors for fourteen semiconductors of the diamond and
  zinc-blende structures},\ }\href {https://doi.org/10.1103/PhysRev.141.789}
  {\bibfield  {journal} {\bibinfo  {journal} {Phys. Rev.}\ }\textbf {\bibinfo
  {volume} {141}},\ \bibinfo {pages} {789} (\bibinfo {year}
  {1966})}\BibitemShut {NoStop}%
\bibitem [{\citenamefont {Chelikowsky}\ and\ \citenamefont
  {Cohen}(1976)}]{Chelikowsky1976}%
  \BibitemOpen
  \bibfield  {author} {\bibinfo {author} {\bibfnamefont {J.~R.}\ \bibnamefont
  {Chelikowsky}}\ and\ \bibinfo {author} {\bibfnamefont {M.~L.}\ \bibnamefont
  {Cohen}},\ }\bibfield  {title} {\bibinfo {title} {Nonlocal pseudopotential
  calculations for the electronic structure of eleven diamond and zinc-blende
  semiconductors},\ }\href {https://doi.org/10.1103/PhysRevB.14.556} {\bibfield
   {journal} {\bibinfo  {journal} {Phys. Rev. B}\ }\textbf {\bibinfo {volume}
  {14}},\ \bibinfo {pages} {556} (\bibinfo {year} {1976})}\BibitemShut
  {NoStop}%
\bibitem [{\citenamefont {Cohen}\ and\ \citenamefont
  {Chelikowsky}(1988)}]{CohenChelikowsky1988}%
  \BibitemOpen
  \bibfield  {author} {\bibinfo {author} {\bibfnamefont {M.~L.}\ \bibnamefont
  {Cohen}}\ and\ \bibinfo {author} {\bibfnamefont {J.~R.}\ \bibnamefont
  {Chelikowsky}},\ }\href@noop {} {\emph {\bibinfo {title} {Electronic
  Structure and Optical Properties of Semiconductors}}}\ (\bibinfo  {publisher}
  {Springer},\ \bibinfo {address} {Berlin},\ \bibinfo {year}
  {1988})\BibitemShut {NoStop}%
\bibitem [{\citenamefont {Wang}\ and\ \citenamefont {Zunger}(1994)}]{Wang1994}%
  \BibitemOpen
  \bibfield  {author} {\bibinfo {author} {\bibfnamefont {L.-W.}\ \bibnamefont
  {Wang}}\ and\ \bibinfo {author} {\bibfnamefont {A.}~\bibnamefont {Zunger}},\
  }\bibfield  {title} {\bibinfo {title} {Solving {S}chr\"odinger's equation
  around a desired energy: Application to silicon quantum dots},\ }\href
  {https://doi.org/10.1063/1.466486} {\bibfield  {journal} {\bibinfo  {journal}
  {J. Chem. Phys.}\ }\textbf {\bibinfo {volume} {100}},\ \bibinfo {pages}
  {2394} (\bibinfo {year} {1994})}\BibitemShut {NoStop}%
\bibitem [{\citenamefont {Wang}\ and\ \citenamefont {Zunger}(1995)}]{Wang1995}%
  \BibitemOpen
  \bibfield  {author} {\bibinfo {author} {\bibfnamefont {L.-W.}\ \bibnamefont
  {Wang}}\ and\ \bibinfo {author} {\bibfnamefont {A.}~\bibnamefont {Zunger}},\
  }\bibfield  {title} {\bibinfo {title} {Local-density-derived semiempirical
  pseudopotentials},\ }\href {https://doi.org/10.1103/PhysRevB.51.17398}
  {\bibfield  {journal} {\bibinfo  {journal} {Phys. Rev. B}\ }\textbf {\bibinfo
  {volume} {51}},\ \bibinfo {pages} {17398} (\bibinfo {year}
  {1995})}\BibitemShut {NoStop}%
\bibitem [{\citenamefont {Hohenberg}\ and\ \citenamefont
  {Kohn}(1964)}]{Hohenberg1964}%
  \BibitemOpen
  \bibfield  {author} {\bibinfo {author} {\bibfnamefont {P.}~\bibnamefont
  {Hohenberg}}\ and\ \bibinfo {author} {\bibfnamefont {W.}~\bibnamefont
  {Kohn}},\ }\bibfield  {title} {\bibinfo {title} {Inhomogeneous electron
  gas},\ }\href {https://doi.org/10.1103/PhysRev.136.B864} {\bibfield
  {journal} {\bibinfo  {journal} {Phys. Rev.}\ }\textbf {\bibinfo {volume}
  {136}},\ \bibinfo {pages} {B864} (\bibinfo {year} {1964})}\BibitemShut
  {NoStop}%
\bibitem [{\citenamefont {Kohn}\ and\ \citenamefont {Sham}(1965)}]{Kohn1965}%
  \BibitemOpen
  \bibfield  {author} {\bibinfo {author} {\bibfnamefont {W.}~\bibnamefont
  {Kohn}}\ and\ \bibinfo {author} {\bibfnamefont {L.~J.}\ \bibnamefont
  {Sham}},\ }\bibfield  {title} {\bibinfo {title} {Self-consistent equations
  including exchange and correlation effects},\ }\href
  {https://doi.org/10.1103/PhysRev.140.A1133} {\bibfield  {journal} {\bibinfo
  {journal} {Phys. Rev.}\ }\textbf {\bibinfo {volume} {140}},\ \bibinfo {pages}
  {A1133} (\bibinfo {year} {1965})}\BibitemShut {NoStop}%
\bibitem [{\citenamefont {Bl\"ochl}(1994)}]{Blochl1994}%
  \BibitemOpen
  \bibfield  {author} {\bibinfo {author} {\bibfnamefont {P.~E.}\ \bibnamefont
  {Bl\"ochl}},\ }\bibfield  {title} {\bibinfo {title} {Projector augmented-wave
  method},\ }\href {https://doi.org/10.1103/PhysRevB.50.17953} {\bibfield
  {journal} {\bibinfo  {journal} {Phys. Rev. B}\ }\textbf {\bibinfo {volume}
  {50}},\ \bibinfo {pages} {17953} (\bibinfo {year} {1994})}\BibitemShut
  {NoStop}%
\bibitem [{\citenamefont {Kresse}\ and\ \citenamefont
  {Joubert}(1999)}]{Kresse1999}%
  \BibitemOpen
  \bibfield  {author} {\bibinfo {author} {\bibfnamefont {G.}~\bibnamefont
  {Kresse}}\ and\ \bibinfo {author} {\bibfnamefont {D.}~\bibnamefont
  {Joubert}},\ }\bibfield  {title} {\bibinfo {title} {From ultrasoft
  pseudopotentials to the projector augmented-wave method},\ }\href
  {https://doi.org/10.1103/PhysRevB.59.1758} {\bibfield  {journal} {\bibinfo
  {journal} {Phys. Rev. B}\ }\textbf {\bibinfo {volume} {59}},\ \bibinfo
  {pages} {1758} (\bibinfo {year} {1999})}\BibitemShut {NoStop}%
\bibitem [{\citenamefont {Slater}\ and\ \citenamefont
  {Koster}(1954)}]{Slater1954}%
  \BibitemOpen
  \bibfield  {author} {\bibinfo {author} {\bibfnamefont {J.~C.}\ \bibnamefont
  {Slater}}\ and\ \bibinfo {author} {\bibfnamefont {G.~F.}\ \bibnamefont
  {Koster}},\ }\bibfield  {title} {\bibinfo {title} {Simplified {LCAO} method
  for the periodic potential problem},\ }\href
  {https://doi.org/10.1103/PhysRev.94.1498} {\bibfield  {journal} {\bibinfo
  {journal} {Phys. Rev.}\ }\textbf {\bibinfo {volume} {94}},\ \bibinfo {pages}
  {1498} (\bibinfo {year} {1954})}\BibitemShut {NoStop}%
\bibitem [{\citenamefont {Vogl}\ \emph {et~al.}(1983)\citenamefont {Vogl},
  \citenamefont {Hjalmarson},\ and\ \citenamefont {Dow}}]{Vogl1983}%
  \BibitemOpen
  \bibfield  {author} {\bibinfo {author} {\bibfnamefont {P.}~\bibnamefont
  {Vogl}}, \bibinfo {author} {\bibfnamefont {H.~P.}\ \bibnamefont
  {Hjalmarson}},\ and\ \bibinfo {author} {\bibfnamefont {J.~D.}\ \bibnamefont
  {Dow}},\ }\bibfield  {title} {\bibinfo {title} {A semi-empirical
  tight-binding theory of the electronic structure of semiconductors},\ }\href
  {https://doi.org/10.1016/0022-3697(83)90064-1} {\bibfield  {journal}
  {\bibinfo  {journal} {J. Phys. Chem. Solids}\ }\textbf {\bibinfo {volume}
  {44}},\ \bibinfo {pages} {365} (\bibinfo {year} {1983})}\BibitemShut
  {NoStop}%
\bibitem [{\citenamefont {Jancu}\ \emph {et~al.}(1998)\citenamefont {Jancu},
  \citenamefont {Scholz}, \citenamefont {Beltram},\ and\ \citenamefont
  {Bassani}}]{Jancu1998}%
  \BibitemOpen
  \bibfield  {author} {\bibinfo {author} {\bibfnamefont {J.-M.}\ \bibnamefont
  {Jancu}}, \bibinfo {author} {\bibfnamefont {R.}~\bibnamefont {Scholz}},
  \bibinfo {author} {\bibfnamefont {F.}~\bibnamefont {Beltram}},\ and\ \bibinfo
  {author} {\bibfnamefont {F.}~\bibnamefont {Bassani}},\ }\bibfield  {title}
  {\bibinfo {title} {Empirical $spds^*$ tight-binding calculation for cubic
  semiconductors: General method and material parameters},\ }\href
  {https://doi.org/10.1103/PhysRevB.57.6493} {\bibfield  {journal} {\bibinfo
  {journal} {Phys. Rev. B}\ }\textbf {\bibinfo {volume} {57}},\ \bibinfo
  {pages} {6493} (\bibinfo {year} {1998})}\BibitemShut {NoStop}%
\bibitem [{\citenamefont {Batzner}\ \emph {et~al.}(2022)\citenamefont
  {Batzner}, \citenamefont {Musaelian}, \citenamefont {Sun}, \citenamefont
  {Geiger}, \citenamefont {Mailoa}, \citenamefont {Kornbluth}, \citenamefont
  {Molinari}, \citenamefont {Smidt},\ and\ \citenamefont
  {Kozinsky}}]{Batzner2022}%
  \BibitemOpen
  \bibfield  {author} {\bibinfo {author} {\bibfnamefont {S.}~\bibnamefont
  {Batzner}}, \bibinfo {author} {\bibfnamefont {A.}~\bibnamefont {Musaelian}},
  \bibinfo {author} {\bibfnamefont {L.}~\bibnamefont {Sun}}, \bibinfo {author}
  {\bibfnamefont {M.}~\bibnamefont {Geiger}}, \bibinfo {author} {\bibfnamefont
  {J.~P.}\ \bibnamefont {Mailoa}}, \bibinfo {author} {\bibfnamefont
  {M.}~\bibnamefont {Kornbluth}}, \bibinfo {author} {\bibfnamefont
  {N.}~\bibnamefont {Molinari}}, \bibinfo {author} {\bibfnamefont {T.~E.}\
  \bibnamefont {Smidt}},\ and\ \bibinfo {author} {\bibfnamefont
  {B.}~\bibnamefont {Kozinsky}},\ }\bibfield  {title} {\bibinfo {title}
  {E(3)-equivariant graph neural networks for data-efficient and accurate
  interatomic potentials},\ }\href {https://doi.org/10.1038/s41467-022-29939-5}
  {\bibfield  {journal} {\bibinfo  {journal} {Nat. Commun.}\ }\textbf {\bibinfo
  {volume} {13}},\ \bibinfo {pages} {2453} (\bibinfo {year}
  {2022})}\BibitemShut {NoStop}%
\bibitem [{\citenamefont {Kim}\ and\ \citenamefont {Son}(2024)}]{KimSon2024}%
  \BibitemOpen
  \bibfield  {author} {\bibinfo {author} {\bibfnamefont {R.}~\bibnamefont
  {Kim}}\ and\ \bibinfo {author} {\bibfnamefont {Y.-W.}\ \bibnamefont {Son}},\
  }\bibfield  {title} {\bibinfo {title} {Transferable empirical
  pseudopotentials from machine learning},\ }\href
  {https://doi.org/10.1103/PhysRevB.109.045153} {\bibfield  {journal} {\bibinfo
   {journal} {Phys. Rev. B}\ }\textbf {\bibinfo {volume} {109}},\ \bibinfo
  {pages} {045153} (\bibinfo {year} {2024})}\BibitemShut {NoStop}%
\bibitem [{\citenamefont {Lin}\ \emph {et~al.}(2025)\citenamefont {Lin},
  \citenamefont {Coley-O'Rourke},\ and\ \citenamefont {Rabani}}]{Lin2025}%
  \BibitemOpen
  \bibfield  {author} {\bibinfo {author} {\bibfnamefont {K.}~\bibnamefont
  {Lin}}, \bibinfo {author} {\bibfnamefont {M.~J.}\ \bibnamefont
  {Coley-O'Rourke}},\ and\ \bibinfo {author} {\bibfnamefont {E.}~\bibnamefont
  {Rabani}},\ }\bibfield  {title} {\bibinfo {title} {Deep-learning atomistic
  semi-empirical pseudopotential model for nanomaterials},\ }\href
  {https://doi.org/10.1038/s41524-025-01862-5} {\bibfield  {journal} {\bibinfo
  {journal} {npj Comput. Mater.}\ }\textbf {\bibinfo {volume} {11}},\ \bibinfo
  {pages} {381} (\bibinfo {year} {2025})}\BibitemShut {NoStop}%
\bibitem [{\citenamefont {Li}\ \emph {et~al.}(2022)\citenamefont {Li},
  \citenamefont {Wang}, \citenamefont {Zou}, \citenamefont {Ye}, \citenamefont
  {Xu}, \citenamefont {Gong}, \citenamefont {Duan},\ and\ \citenamefont
  {Xu}}]{Li2022DeepH}%
  \BibitemOpen
  \bibfield  {author} {\bibinfo {author} {\bibfnamefont {H.}~\bibnamefont
  {Li}}, \bibinfo {author} {\bibfnamefont {Z.}~\bibnamefont {Wang}}, \bibinfo
  {author} {\bibfnamefont {N.}~\bibnamefont {Zou}}, \bibinfo {author}
  {\bibfnamefont {M.}~\bibnamefont {Ye}}, \bibinfo {author} {\bibfnamefont
  {R.}~\bibnamefont {Xu}}, \bibinfo {author} {\bibfnamefont {X.}~\bibnamefont
  {Gong}}, \bibinfo {author} {\bibfnamefont {W.}~\bibnamefont {Duan}},\ and\
  \bibinfo {author} {\bibfnamefont {Y.}~\bibnamefont {Xu}},\ }\bibfield
  {title} {\bibinfo {title} {Deep-learning density functional theory
  {H}amiltonian for efficient ab initio electronic-structure calculation},\
  }\href {https://doi.org/10.1038/s43588-022-00265-6} {\bibfield  {journal}
  {\bibinfo  {journal} {Nat. Comput. Sci.}\ }\textbf {\bibinfo {volume} {2}},\
  \bibinfo {pages} {367} (\bibinfo {year} {2022})}\BibitemShut {NoStop}%
\bibitem [{\citenamefont {Zhong}\ \emph {et~al.}(2023)\citenamefont {Zhong},
  \citenamefont {Yu}, \citenamefont {Su}, \citenamefont {Gong},\ and\
  \citenamefont {Xiang}}]{Zhong2023HamGNN}%
  \BibitemOpen
  \bibfield  {author} {\bibinfo {author} {\bibfnamefont {Y.}~\bibnamefont
  {Zhong}}, \bibinfo {author} {\bibfnamefont {H.}~\bibnamefont {Yu}}, \bibinfo
  {author} {\bibfnamefont {M.}~\bibnamefont {Su}}, \bibinfo {author}
  {\bibfnamefont {X.}~\bibnamefont {Gong}},\ and\ \bibinfo {author}
  {\bibfnamefont {H.}~\bibnamefont {Xiang}},\ }\bibfield  {title} {\bibinfo
  {title} {Transferable equivariant graph neural networks for the
  {H}amiltonians of molecules and solids},\ }\href
  {https://doi.org/10.1038/s41524-023-01130-4} {\bibfield  {journal} {\bibinfo
  {journal} {npj Comput. Mater.}\ }\textbf {\bibinfo {volume} {9}},\ \bibinfo
  {pages} {182} (\bibinfo {year} {2023})}\BibitemShut {NoStop}%
\bibitem [{\citenamefont {Gu}\ \emph {et~al.}(2024)\citenamefont {Gu} \emph
  {et~al.}}]{Gu2024DeePTB}%
  \BibitemOpen
  \bibfield  {author} {\bibinfo {author} {\bibfnamefont {Q.}~\bibnamefont {Gu}}
  \emph {et~al.},\ }\bibfield  {title} {\bibinfo {title} {Deep learning
  tight-binding approach for large-scale electronic simulations at finite
  temperatures with ab initio accuracy},\ }\href
  {https://doi.org/10.1038/s41467-024-51006-4} {\bibfield  {journal} {\bibinfo
  {journal} {Nat. Commun.}\ }\textbf {\bibinfo {volume} {15}},\ \bibinfo
  {pages} {6772} (\bibinfo {year} {2024})}\BibitemShut {NoStop}%
\bibitem [{\citenamefont {Zhong}\ \emph {et~al.}(2026)\citenamefont {Zhong},
  \citenamefont {Li}, \citenamefont {Gong},\ and\ \citenamefont
  {Xiang}}]{Zhong2026LR}%
  \BibitemOpen
  \bibfield  {author} {\bibinfo {author} {\bibfnamefont {Y.}~\bibnamefont
  {Zhong}}, \bibinfo {author} {\bibfnamefont {X.}~\bibnamefont {Li}}, \bibinfo
  {author} {\bibfnamefont {X.}~\bibnamefont {Gong}},\ and\ \bibinfo {author}
  {\bibfnamefont {H.}~\bibnamefont {Xiang}},\ }\href@noop {} {\bibinfo {title}
  {Physics-informed long-range {C}oulomb correction for machine-learning
  {H}amiltonians}} (\bibinfo {year} {2026}),\ \Eprint
  {https://arxiv.org/abs/2603.20007} {arXiv:2603.20007 [cond-mat.mtrl-sci]}
  \BibitemShut {NoStop}%
\bibitem [{\citenamefont {Zhang}\ \emph {et~al.}(2026)\citenamefont {Zhang},
  \citenamefont {Cui}, \citenamefont {Zhou}, \citenamefont {Bai}, \citenamefont
  {Zhang}, \citenamefont {Rossi}, \citenamefont {Su}, \citenamefont {Ouyang},\
  and\ \citenamefont {Heng}}]{Zhang2026Ewald}%
  \BibitemOpen
  \bibfield  {author} {\bibinfo {author} {\bibfnamefont {L.}~\bibnamefont
  {Zhang}}, \bibinfo {author} {\bibfnamefont {T.}~\bibnamefont {Cui}}, \bibinfo
  {author} {\bibfnamefont {D.}~\bibnamefont {Zhou}}, \bibinfo {author}
  {\bibfnamefont {L.}~\bibnamefont {Bai}}, \bibinfo {author} {\bibfnamefont
  {S.}~\bibnamefont {Zhang}}, \bibinfo {author} {\bibfnamefont
  {L.}~\bibnamefont {Rossi}}, \bibinfo {author} {\bibfnamefont
  {M.}~\bibnamefont {Su}}, \bibinfo {author} {\bibfnamefont {W.}~\bibnamefont
  {Ouyang}},\ and\ \bibinfo {author} {\bibfnamefont {P.-A.}\ \bibnamefont
  {Heng}},\ }\href@noop {} {\bibinfo {title} {An {SO(3)}-equivariant
  reciprocal-space neural potential for long-range interactions}} (\bibinfo
  {year} {2026}),\ \Eprint {https://arxiv.org/abs/2603.18389}
  {arXiv:2603.18389} \BibitemShut {NoStop}%
\bibitem [{\citenamefont {Hauge}\ \emph {et~al.}(2015)\citenamefont {Hauge},
  \citenamefont {Verheijen}, \citenamefont {Conesa-Boj}, \citenamefont
  {Etzelstorfer}, \citenamefont {Watzinger}, \citenamefont {Kriegner},
  \citenamefont {Zardo}, \citenamefont {Fasolato}, \citenamefont {Capitani},
  \citenamefont {Postorino}, \citenamefont {K\"olling}, \citenamefont {Li},
  \citenamefont {Assali}, \citenamefont {Stangl},\ and\ \citenamefont
  {Bakkers}}]{Hauge2015}%
  \BibitemOpen
  \bibfield  {author} {\bibinfo {author} {\bibfnamefont {H.~I.~T.}\
  \bibnamefont {Hauge}}, \bibinfo {author} {\bibfnamefont {M.~A.}\ \bibnamefont
  {Verheijen}}, \bibinfo {author} {\bibfnamefont {S.}~\bibnamefont
  {Conesa-Boj}}, \bibinfo {author} {\bibfnamefont {T.}~\bibnamefont
  {Etzelstorfer}}, \bibinfo {author} {\bibfnamefont {M.}~\bibnamefont
  {Watzinger}}, \bibinfo {author} {\bibfnamefont {D.}~\bibnamefont {Kriegner}},
  \bibinfo {author} {\bibfnamefont {I.}~\bibnamefont {Zardo}}, \bibinfo
  {author} {\bibfnamefont {C.}~\bibnamefont {Fasolato}}, \bibinfo {author}
  {\bibfnamefont {F.}~\bibnamefont {Capitani}}, \bibinfo {author}
  {\bibfnamefont {P.}~\bibnamefont {Postorino}}, \bibinfo {author}
  {\bibfnamefont {S.}~\bibnamefont {K\"olling}}, \bibinfo {author}
  {\bibfnamefont {A.}~\bibnamefont {Li}}, \bibinfo {author} {\bibfnamefont
  {S.}~\bibnamefont {Assali}}, \bibinfo {author} {\bibfnamefont
  {J.}~\bibnamefont {Stangl}},\ and\ \bibinfo {author} {\bibfnamefont {E.~P.
  A.~M.}\ \bibnamefont {Bakkers}},\ }\bibfield  {title} {\bibinfo {title}
  {Hexagonal silicon realized},\ }\href
  {https://doi.org/10.1021/acs.nanolett.5b01939} {\bibfield  {journal}
  {\bibinfo  {journal} {Nano Lett.}\ }\textbf {\bibinfo {volume} {15}},\
  \bibinfo {pages} {5855} (\bibinfo {year} {2015})}\BibitemShut {NoStop}%
\bibitem [{\citenamefont {Fadaly}\ \emph {et~al.}(2020)\citenamefont {Fadaly}
  \emph {et~al.}}]{Fadaly2020}%
  \BibitemOpen
  \bibfield  {author} {\bibinfo {author} {\bibfnamefont {E.~M.~T.}\
  \bibnamefont {Fadaly}} \emph {et~al.},\ }\bibfield  {title} {\bibinfo {title}
  {Direct-bandgap emission from hexagonal {Ge} and {SiGe} alloys},\ }\href
  {https://doi.org/10.1038/s41586-020-2150-y} {\bibfield  {journal} {\bibinfo
  {journal} {Nature}\ }\textbf {\bibinfo {volume} {580}},\ \bibinfo {pages}
  {205} (\bibinfo {year} {2020})}\BibitemShut {NoStop}%
\bibitem [{\citenamefont {Gryko}\ \emph {et~al.}(2000)\citenamefont {Gryko},
  \citenamefont {McMillan}, \citenamefont {Marzke}, \citenamefont
  {Ramachandran}, \citenamefont {Patton}, \citenamefont {Deb},\ and\
  \citenamefont {Sankey}}]{Gryko2000}%
  \BibitemOpen
  \bibfield  {author} {\bibinfo {author} {\bibfnamefont {J.}~\bibnamefont
  {Gryko}}, \bibinfo {author} {\bibfnamefont {P.~F.}\ \bibnamefont {McMillan}},
  \bibinfo {author} {\bibfnamefont {R.~F.}\ \bibnamefont {Marzke}}, \bibinfo
  {author} {\bibfnamefont {G.~K.}\ \bibnamefont {Ramachandran}}, \bibinfo
  {author} {\bibfnamefont {D.}~\bibnamefont {Patton}}, \bibinfo {author}
  {\bibfnamefont {S.~K.}\ \bibnamefont {Deb}},\ and\ \bibinfo {author}
  {\bibfnamefont {O.~F.}\ \bibnamefont {Sankey}},\ }\bibfield  {title}
  {\bibinfo {title} {Low-density framework form of crystalline silicon with a
  wide optical band gap},\ }\href {https://doi.org/10.1103/PhysRevB.62.R7707}
  {\bibfield  {journal} {\bibinfo  {journal} {Phys. Rev. B}\ }\textbf {\bibinfo
  {volume} {62}},\ \bibinfo {pages} {R7707} (\bibinfo {year}
  {2000})}\BibitemShut {NoStop}%
\bibitem [{\citenamefont {Kim}\ \emph {et~al.}(2015)\citenamefont {Kim},
  \citenamefont {Stefanoski}, \citenamefont {Kurakevych},\ and\ \citenamefont
  {Strobel}}]{Kim2015Si24}%
  \BibitemOpen
  \bibfield  {author} {\bibinfo {author} {\bibfnamefont {D.~Y.}\ \bibnamefont
  {Kim}}, \bibinfo {author} {\bibfnamefont {S.}~\bibnamefont {Stefanoski}},
  \bibinfo {author} {\bibfnamefont {O.~O.}\ \bibnamefont {Kurakevych}},\ and\
  \bibinfo {author} {\bibfnamefont {T.~A.}\ \bibnamefont {Strobel}},\
  }\bibfield  {title} {\bibinfo {title} {Synthesis of an open-framework
  allotrope of silicon},\ }\href {https://doi.org/10.1038/nmat4140} {\bibfield
  {journal} {\bibinfo  {journal} {Nat. Mater.}\ }\textbf {\bibinfo {volume}
  {14}},\ \bibinfo {pages} {169} (\bibinfo {year} {2015})}\BibitemShut
  {NoStop}%
\bibitem [{\citenamefont {Mujica}\ \emph {et~al.}(2003)\citenamefont {Mujica},
  \citenamefont {Rubio}, \citenamefont {Mu\~noz},\ and\ \citenamefont
  {Needs}}]{Mujica2003}%
  \BibitemOpen
  \bibfield  {author} {\bibinfo {author} {\bibfnamefont {A.}~\bibnamefont
  {Mujica}}, \bibinfo {author} {\bibfnamefont {A.}~\bibnamefont {Rubio}},
  \bibinfo {author} {\bibfnamefont {A.}~\bibnamefont {Mu\~noz}},\ and\ \bibinfo
  {author} {\bibfnamefont {R.~J.}\ \bibnamefont {Needs}},\ }\bibfield  {title}
  {\bibinfo {title} {High-pressure phases of group-{IV}, {III}--{V}, and
  {II}--{VI} compounds},\ }\href {https://doi.org/10.1103/RevModPhys.75.863}
  {\bibfield  {journal} {\bibinfo  {journal} {Rev. Mod. Phys.}\ }\textbf
  {\bibinfo {volume} {75}},\ \bibinfo {pages} {863} (\bibinfo {year}
  {2003})}\BibitemShut {NoStop}%
\bibitem [{\citenamefont {Perdew}\ \emph {et~al.}(1996)\citenamefont {Perdew},
  \citenamefont {Burke},\ and\ \citenamefont {Ernzerhof}}]{Perdew1996}%
  \BibitemOpen
  \bibfield  {author} {\bibinfo {author} {\bibfnamefont {J.~P.}\ \bibnamefont
  {Perdew}}, \bibinfo {author} {\bibfnamefont {K.}~\bibnamefont {Burke}},\ and\
  \bibinfo {author} {\bibfnamefont {M.}~\bibnamefont {Ernzerhof}},\ }\bibfield
  {title} {\bibinfo {title} {Generalized gradient approximation made simple},\
  }\href {https://doi.org/10.1103/PhysRevLett.77.3865} {\bibfield  {journal}
  {\bibinfo  {journal} {Phys. Rev. Lett.}\ }\textbf {\bibinfo {volume} {77}},\
  \bibinfo {pages} {3865} (\bibinfo {year} {1996})}\BibitemShut {NoStop}%
\bibitem [{\citenamefont {Jacobs}\ \emph {et~al.}(1991)\citenamefont {Jacobs},
  \citenamefont {Jordan}, \citenamefont {Nowlan},\ and\ \citenamefont
  {Hinton}}]{Jacobs1991}%
  \BibitemOpen
  \bibfield  {author} {\bibinfo {author} {\bibfnamefont {R.~A.}\ \bibnamefont
  {Jacobs}}, \bibinfo {author} {\bibfnamefont {M.~I.}\ \bibnamefont {Jordan}},
  \bibinfo {author} {\bibfnamefont {S.~J.}\ \bibnamefont {Nowlan}},\ and\
  \bibinfo {author} {\bibfnamefont {G.~E.}\ \bibnamefont {Hinton}},\ }\bibfield
   {title} {\bibinfo {title} {Adaptive mixtures of local experts},\ }\href
  {https://doi.org/10.1162/neco.1991.3.1.79} {\bibfield  {journal} {\bibinfo
  {journal} {Neural Comput.}\ }\textbf {\bibinfo {volume} {3}},\ \bibinfo
  {pages} {79} (\bibinfo {year} {1991})}\BibitemShut {NoStop}%
\bibitem [{\citenamefont {Shazeer}\ \emph {et~al.}(2017)\citenamefont
  {Shazeer}, \citenamefont {Mirhoseini}, \citenamefont {Maziarz}, \citenamefont
  {Davis}, \citenamefont {Le}, \citenamefont {Hinton},\ and\ \citenamefont
  {Dean}}]{Shazeer2017}%
  \BibitemOpen
  \bibfield  {author} {\bibinfo {author} {\bibfnamefont {N.}~\bibnamefont
  {Shazeer}}, \bibinfo {author} {\bibfnamefont {A.}~\bibnamefont {Mirhoseini}},
  \bibinfo {author} {\bibfnamefont {K.}~\bibnamefont {Maziarz}}, \bibinfo
  {author} {\bibfnamefont {A.}~\bibnamefont {Davis}}, \bibinfo {author}
  {\bibfnamefont {Q.}~\bibnamefont {Le}}, \bibinfo {author} {\bibfnamefont
  {G.}~\bibnamefont {Hinton}},\ and\ \bibinfo {author} {\bibfnamefont
  {J.}~\bibnamefont {Dean}},\ }\bibfield  {title} {\bibinfo {title}
  {Outrageously large neural networks: The sparsely-gated mixture-of-experts
  layer},\ }in\ \href@noop {} {\emph {\bibinfo {booktitle} {International
  Conference on Learning Representations (ICLR)}}}\ (\bibinfo {year} {2017})\
  \Eprint {https://arxiv.org/abs/1701.06538} {arXiv:1701.06538} \BibitemShut
  {NoStop}%
\bibitem [{\citenamefont {Kleinman}\ and\ \citenamefont
  {Bylander}(1982)}]{Kleinman1982}%
  \BibitemOpen
  \bibfield  {author} {\bibinfo {author} {\bibfnamefont {L.}~\bibnamefont
  {Kleinman}}\ and\ \bibinfo {author} {\bibfnamefont {D.~M.}\ \bibnamefont
  {Bylander}},\ }\bibfield  {title} {\bibinfo {title} {Efficacious form for
  model pseudopotentials},\ }\href
  {https://doi.org/10.1103/PhysRevLett.48.1425} {\bibfield  {journal} {\bibinfo
   {journal} {Phys. Rev. Lett.}\ }\textbf {\bibinfo {volume} {48}},\ \bibinfo
  {pages} {1425} (\bibinfo {year} {1982})}\BibitemShut {NoStop}%
\bibitem [{\citenamefont {Giannozzi}\ \emph {et~al.}(2009)\citenamefont
  {Giannozzi} \emph {et~al.}}]{Giannozzi2009}%
  \BibitemOpen
  \bibfield  {author} {\bibinfo {author} {\bibfnamefont {P.}~\bibnamefont
  {Giannozzi}} \emph {et~al.},\ }\bibfield  {title} {\bibinfo {title} {{QUANTUM
  ESPRESSO}: a modular and open-source software project for quantum simulations
  of materials},\ }\href {https://doi.org/10.1088/0953-8984/21/39/395502}
  {\bibfield  {journal} {\bibinfo  {journal} {J. Phys.: Condens. Matter}\
  }\textbf {\bibinfo {volume} {21}},\ \bibinfo {pages} {395502} (\bibinfo
  {year} {2009})}\BibitemShut {NoStop}%
\bibitem [{\citenamefont {Giannozzi}\ \emph {et~al.}(2017)\citenamefont
  {Giannozzi} \emph {et~al.}}]{Giannozzi2017}%
  \BibitemOpen
  \bibfield  {author} {\bibinfo {author} {\bibfnamefont {P.}~\bibnamefont
  {Giannozzi}} \emph {et~al.},\ }\bibfield  {title} {\bibinfo {title} {Advanced
  capabilities for materials modelling with {Quantum ESPRESSO}},\ }\href
  {https://doi.org/10.1088/1361-648X/aa8f79} {\bibfield  {journal} {\bibinfo
  {journal} {J. Phys.: Condens. Matter}\ }\textbf {\bibinfo {volume} {29}},\
  \bibinfo {pages} {465901} (\bibinfo {year} {2017})}\BibitemShut {NoStop}%
\bibitem [{\citenamefont {Dal~Corso}(2014)}]{DalCorso2014}%
  \BibitemOpen
  \bibfield  {author} {\bibinfo {author} {\bibfnamefont {A.}~\bibnamefont
  {Dal~Corso}},\ }\bibfield  {title} {\bibinfo {title} {Pseudopotentials
  periodic table: From {H} to {Pu}},\ }\href
  {https://doi.org/10.1016/j.commatsci.2014.07.043} {\bibfield  {journal}
  {\bibinfo  {journal} {Comput. Mater. Sci.}\ }\textbf {\bibinfo {volume}
  {95}},\ \bibinfo {pages} {337} (\bibinfo {year} {2014})}\BibitemShut
  {NoStop}%
\bibitem [{\citenamefont {Hinuma}\ \emph {et~al.}(2017)\citenamefont {Hinuma},
  \citenamefont {Pizzi}, \citenamefont {Kumagai}, \citenamefont {Oba},\ and\
  \citenamefont {Tanaka}}]{Hinuma2017}%
  \BibitemOpen
  \bibfield  {author} {\bibinfo {author} {\bibfnamefont {Y.}~\bibnamefont
  {Hinuma}}, \bibinfo {author} {\bibfnamefont {G.}~\bibnamefont {Pizzi}},
  \bibinfo {author} {\bibfnamefont {Y.}~\bibnamefont {Kumagai}}, \bibinfo
  {author} {\bibfnamefont {F.}~\bibnamefont {Oba}},\ and\ \bibinfo {author}
  {\bibfnamefont {I.}~\bibnamefont {Tanaka}},\ }\bibfield  {title} {\bibinfo
  {title} {Band structure diagram paths based on crystallography},\ }\href
  {https://doi.org/10.1016/j.commatsci.2016.10.015} {\bibfield  {journal}
  {\bibinfo  {journal} {Comput. Mater. Sci.}\ }\textbf {\bibinfo {volume}
  {128}},\ \bibinfo {pages} {140} (\bibinfo {year} {2017})}\BibitemShut
  {NoStop}%
\bibitem [{\citenamefont {Marzari}\ \emph {et~al.}(1999)\citenamefont
  {Marzari}, \citenamefont {Vanderbilt}, \citenamefont {De~Vita},\ and\
  \citenamefont {Payne}}]{Marzari1999}%
  \BibitemOpen
  \bibfield  {author} {\bibinfo {author} {\bibfnamefont {N.}~\bibnamefont
  {Marzari}}, \bibinfo {author} {\bibfnamefont {D.}~\bibnamefont {Vanderbilt}},
  \bibinfo {author} {\bibfnamefont {A.}~\bibnamefont {De~Vita}},\ and\ \bibinfo
  {author} {\bibfnamefont {M.~C.}\ \bibnamefont {Payne}},\ }\bibfield  {title}
  {\bibinfo {title} {Thermal contraction and disordering of the {Al}(110)
  surface},\ }\href {https://doi.org/10.1103/PhysRevLett.82.3296} {\bibfield
  {journal} {\bibinfo  {journal} {Phys. Rev. Lett.}\ }\textbf {\bibinfo
  {volume} {82}},\ \bibinfo {pages} {3296} (\bibinfo {year}
  {1999})}\BibitemShut {NoStop}%
\bibitem [{\citenamefont {Kirklin}\ \emph {et~al.}(2015)\citenamefont
  {Kirklin}, \citenamefont {Saal}, \citenamefont {Meredig}, \citenamefont
  {Thompson}, \citenamefont {Doak}, \citenamefont {Aykol}, \citenamefont
  {R\"uhl},\ and\ \citenamefont {Wolverton}}]{Kirklin2015}%
  \BibitemOpen
  \bibfield  {author} {\bibinfo {author} {\bibfnamefont {S.}~\bibnamefont
  {Kirklin}}, \bibinfo {author} {\bibfnamefont {J.~E.}\ \bibnamefont {Saal}},
  \bibinfo {author} {\bibfnamefont {B.}~\bibnamefont {Meredig}}, \bibinfo
  {author} {\bibfnamefont {A.}~\bibnamefont {Thompson}}, \bibinfo {author}
  {\bibfnamefont {J.~W.}\ \bibnamefont {Doak}}, \bibinfo {author}
  {\bibfnamefont {M.}~\bibnamefont {Aykol}}, \bibinfo {author} {\bibfnamefont
  {S.}~\bibnamefont {R\"uhl}},\ and\ \bibinfo {author} {\bibfnamefont
  {C.}~\bibnamefont {Wolverton}},\ }\bibfield  {title} {\bibinfo {title} {The
  open quantum materials database ({OQMD}): assessing the accuracy of {DFT}
  formation energies},\ }\href {https://doi.org/10.1038/npjcompumats.2015.10}
  {\bibfield  {journal} {\bibinfo  {journal} {npj Comput. Mater.}\ }\textbf
  {\bibinfo {volume} {1}},\ \bibinfo {pages} {15010} (\bibinfo {year}
  {2015})}\BibitemShut {NoStop}%
\bibitem [{\citenamefont {Hoffmann}\ \emph {et~al.}(2016)\citenamefont
  {Hoffmann}, \citenamefont {Kabanov}, \citenamefont {Golov},\ and\
  \citenamefont {Proserpio}}]{Hoffmann2016}%
  \BibitemOpen
  \bibfield  {author} {\bibinfo {author} {\bibfnamefont {R.}~\bibnamefont
  {Hoffmann}}, \bibinfo {author} {\bibfnamefont {A.~A.}\ \bibnamefont
  {Kabanov}}, \bibinfo {author} {\bibfnamefont {A.~A.}\ \bibnamefont {Golov}},\
  and\ \bibinfo {author} {\bibfnamefont {D.~M.}\ \bibnamefont {Proserpio}},\
  }\bibfield  {title} {\bibinfo {title} {Homo citans and carbon allotropes: For
  an ethics of citation},\ }\href {https://doi.org/10.1002/anie.201600655}
  {\bibfield  {journal} {\bibinfo  {journal} {Angew. Chem. Int. Ed.}\ }\textbf
  {\bibinfo {volume} {55}},\ \bibinfo {pages} {10962} (\bibinfo {year}
  {2016})}\BibitemShut {NoStop}%
\bibitem [{\citenamefont {Amsler}\ \emph {et~al.}(2015)\citenamefont {Amsler},
  \citenamefont {Botti}, \citenamefont {Marques}, \citenamefont {Lenosky},\
  and\ \citenamefont {Goedecker}}]{Amsler2015}%
  \BibitemOpen
  \bibfield  {author} {\bibinfo {author} {\bibfnamefont {M.}~\bibnamefont
  {Amsler}}, \bibinfo {author} {\bibfnamefont {S.}~\bibnamefont {Botti}},
  \bibinfo {author} {\bibfnamefont {M.~A.~L.}\ \bibnamefont {Marques}},
  \bibinfo {author} {\bibfnamefont {T.~J.}\ \bibnamefont {Lenosky}},\ and\
  \bibinfo {author} {\bibfnamefont {S.}~\bibnamefont {Goedecker}},\ }\bibfield
  {title} {\bibinfo {title} {Low-density silicon allotropes for photovoltaic
  applications},\ }\href {https://doi.org/10.1103/PhysRevB.92.014101}
  {\bibfield  {journal} {\bibinfo  {journal} {Phys. Rev. B}\ }\textbf {\bibinfo
  {volume} {92}},\ \bibinfo {pages} {014101} (\bibinfo {year}
  {2015})}\BibitemShut {NoStop}%
\bibitem [{\citenamefont {Curtarolo}\ \emph {et~al.}(2012)\citenamefont
  {Curtarolo} \emph {et~al.}}]{Curtarolo2012}%
  \BibitemOpen
  \bibfield  {author} {\bibinfo {author} {\bibfnamefont {S.}~\bibnamefont
  {Curtarolo}} \emph {et~al.},\ }\bibfield  {title} {\bibinfo {title} {{AFLOW}:
  An automatic framework for high-throughput materials discovery},\ }\href
  {https://doi.org/10.1016/j.commatsci.2012.02.005} {\bibfield  {journal}
  {\bibinfo  {journal} {Comput. Mater. Sci.}\ }\textbf {\bibinfo {volume}
  {58}},\ \bibinfo {pages} {218} (\bibinfo {year} {2012})}\BibitemShut
  {NoStop}%
\bibitem [{\citenamefont {Tersoff}(1988)}]{Tersoff1988}%
  \BibitemOpen
  \bibfield  {author} {\bibinfo {author} {\bibfnamefont {J.}~\bibnamefont
  {Tersoff}},\ }\bibfield  {title} {\bibinfo {title} {New empirical approach
  for the structure and energy of covalent systems},\ }\href
  {https://doi.org/10.1103/PhysRevB.37.6991} {\bibfield  {journal} {\bibinfo
  {journal} {Phys. Rev. B}\ }\textbf {\bibinfo {volume} {37}},\ \bibinfo
  {pages} {6991} (\bibinfo {year} {1988})}\BibitemShut {NoStop}%
\bibitem [{\citenamefont {Loshchilov}\ and\ \citenamefont
  {Hutter}(2019)}]{Loshchilov2019}%
  \BibitemOpen
  \bibfield  {author} {\bibinfo {author} {\bibfnamefont {I.}~\bibnamefont
  {Loshchilov}}\ and\ \bibinfo {author} {\bibfnamefont {F.}~\bibnamefont
  {Hutter}},\ }\bibfield  {title} {\bibinfo {title} {Decoupled weight decay
  regularization},\ }in\ \href {https://doi.org/10.48550/arXiv.1711.05101}
  {\emph {\bibinfo {booktitle} {International Conference on Learning
  Representations}}}\ (\bibinfo {year} {2019})\BibitemShut {NoStop}%
\bibitem [{\citenamefont {Birch}(1947)}]{Birch1947}%
  \BibitemOpen
  \bibfield  {author} {\bibinfo {author} {\bibfnamefont {F.}~\bibnamefont
  {Birch}},\ }\bibfield  {title} {\bibinfo {title} {Finite elastic strain of
  cubic crystals},\ }\href {https://doi.org/10.1103/PhysRev.71.809} {\bibfield
  {journal} {\bibinfo  {journal} {Phys. Rev.}\ }\textbf {\bibinfo {volume}
  {71}},\ \bibinfo {pages} {809} (\bibinfo {year} {1947})}\BibitemShut
  {NoStop}%
\bibitem [{\citenamefont {Togo}\ \emph {et~al.}(2024)\citenamefont {Togo},
  \citenamefont {Shinohara},\ and\ \citenamefont {Tanaka}}]{spglib}%
  \BibitemOpen
  \bibfield  {author} {\bibinfo {author} {\bibfnamefont {A.}~\bibnamefont
  {Togo}}, \bibinfo {author} {\bibfnamefont {K.}~\bibnamefont {Shinohara}},\
  and\ \bibinfo {author} {\bibfnamefont {I.}~\bibnamefont {Tanaka}},\
  }\bibfield  {title} {\bibinfo {title} {Spglib: a software library for crystal
  symmetry search},\ }\href {https://doi.org/10.1080/27660400.2024.2384822}
  {\bibfield  {journal} {\bibinfo  {journal} {Sci. Technol. Adv. Mater.,
  Meth.}\ }\textbf {\bibinfo {volume} {4}},\ \bibinfo {pages} {2384822}
  (\bibinfo {year} {2024})}\BibitemShut {NoStop}%
\bibitem [{\citenamefont {Fischetti}\ and\ \citenamefont
  {Laux}(1996)}]{Fischetti1996}%
  \BibitemOpen
  \bibfield  {author} {\bibinfo {author} {\bibfnamefont {M.~V.}\ \bibnamefont
  {Fischetti}}\ and\ \bibinfo {author} {\bibfnamefont {S.~E.}\ \bibnamefont
  {Laux}},\ }\bibfield  {title} {\bibinfo {title} {Band structure, deformation
  potentials, and carrier mobility in strained {Si}, {Ge}, and {SiGe} alloys},\
  }\href {https://doi.org/10.1063/1.363052} {\bibfield  {journal} {\bibinfo
  {journal} {J. Appl. Phys.}\ }\textbf {\bibinfo {volume} {80}},\ \bibinfo
  {pages} {2234} (\bibinfo {year} {1996})}\BibitemShut {NoStop}%
\bibitem [{\citenamefont {Sun}\ \emph {et~al.}(2007)\citenamefont {Sun},
  \citenamefont {Thompson},\ and\ \citenamefont {Nishida}}]{Sun2007}%
  \BibitemOpen
  \bibfield  {author} {\bibinfo {author} {\bibfnamefont {Y.}~\bibnamefont
  {Sun}}, \bibinfo {author} {\bibfnamefont {S.~E.}\ \bibnamefont {Thompson}},\
  and\ \bibinfo {author} {\bibfnamefont {T.}~\bibnamefont {Nishida}},\
  }\bibfield  {title} {\bibinfo {title} {Physics of strain effects in
  semiconductors and metal-oxide-semiconductor field-effect transistors},\
  }\href {https://doi.org/10.1063/1.2730561} {\bibfield  {journal} {\bibinfo
  {journal} {J. Appl. Phys.}\ }\textbf {\bibinfo {volume} {101}},\ \bibinfo
  {pages} {104503} (\bibinfo {year} {2007})}\BibitemShut {NoStop}%
\bibitem [{\citenamefont {Harrison}(1966)}]{Harrison1966}%
  \BibitemOpen
  \bibfield  {author} {\bibinfo {author} {\bibfnamefont {W.~A.}\ \bibnamefont
  {Harrison}},\ }\href@noop {} {\emph {\bibinfo {title} {Pseudopotentials in
  the Theory of Metals}}}\ (\bibinfo  {publisher} {W. A. Benjamin},\ \bibinfo
  {address} {New York},\ \bibinfo {year} {1966})\BibitemShut {NoStop}%
\bibitem [{\citenamefont {Perdew}\ and\ \citenamefont
  {Levy}(1983)}]{Perdew1983}%
  \BibitemOpen
  \bibfield  {author} {\bibinfo {author} {\bibfnamefont {J.~P.}\ \bibnamefont
  {Perdew}}\ and\ \bibinfo {author} {\bibfnamefont {M.}~\bibnamefont {Levy}},\
  }\bibfield  {title} {\bibinfo {title} {Physical content of the exact
  {K}ohn-{S}ham orbital energies: Band gaps and derivative discontinuities},\
  }\href {https://doi.org/10.1103/PhysRevLett.51.1884} {\bibfield  {journal}
  {\bibinfo  {journal} {Phys. Rev. Lett.}\ }\textbf {\bibinfo {volume} {51}},\
  \bibinfo {pages} {1884} (\bibinfo {year} {1983})}\BibitemShut {NoStop}%
\bibitem [{\citenamefont {Hybertsen}\ and\ \citenamefont
  {Louie}(1986)}]{Hybertsen1986}%
  \BibitemOpen
  \bibfield  {author} {\bibinfo {author} {\bibfnamefont {M.~S.}\ \bibnamefont
  {Hybertsen}}\ and\ \bibinfo {author} {\bibfnamefont {S.~G.}\ \bibnamefont
  {Louie}},\ }\bibfield  {title} {\bibinfo {title} {Electron correlation in
  semiconductors and insulators: Band gaps and quasiparticle energies},\ }\href
  {https://doi.org/10.1103/PhysRevB.34.5390} {\bibfield  {journal} {\bibinfo
  {journal} {Phys. Rev. B}\ }\textbf {\bibinfo {volume} {34}},\ \bibinfo
  {pages} {5390} (\bibinfo {year} {1986})}\BibitemShut {NoStop}%
\end{thebibliography}%

\clearpage
\onecolumngrid
\appendix

\section{Supplementary band-structure figures}
\label{app:history}
\label{app:seen}
\label{app:unseen}
\label{app:overlays}

\begin{figure}[!htbp]
\includegraphics[width=0.9\textwidth]{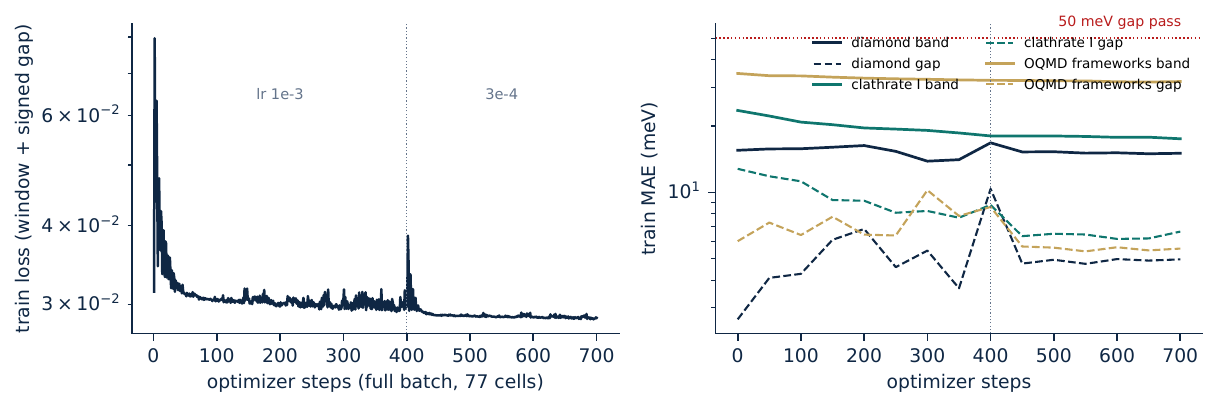}
\caption{\label{fig:history}Training history of EpNet-Si: the final two learning-rate stages (700 steps) on its 77 cells, started from the 66-cell precursor. Left: training loss (dotted line: learning-rate drop). Right: band (solid) and gap (dashed) MAE of monitored training families; the OQMD frameworks keep the highest band error (32 against 15--17\,meV).}
\end{figure}

\begin{figure}[!htbp]
\includegraphics[width=\textwidth]{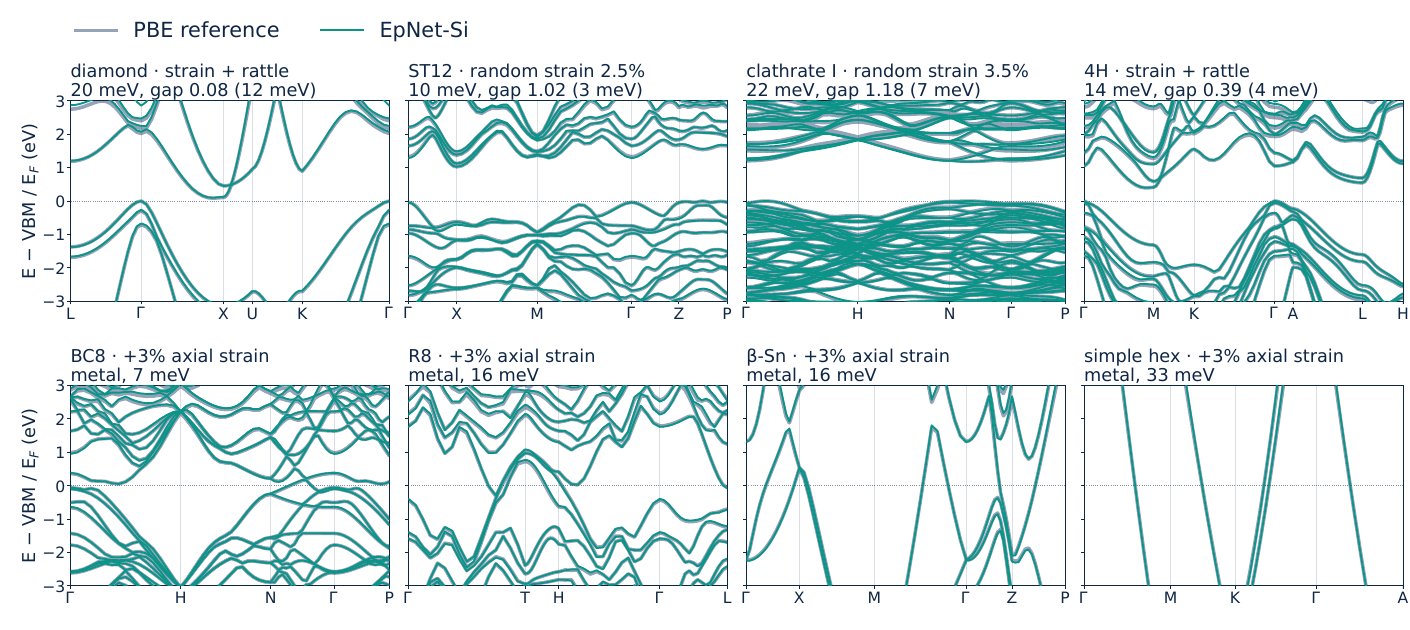}
\caption{\label{fig:known}Distortions of seen structures: one test cell for each of eight selected trained structure types. PBE (grey) and EpNet-Si (teal). Titles give the distortion, the PBE gap (eV), and the band and gap errors (meV).}
\end{figure}

\begin{figure}[!htbp]
\includegraphics[width=\textwidth]{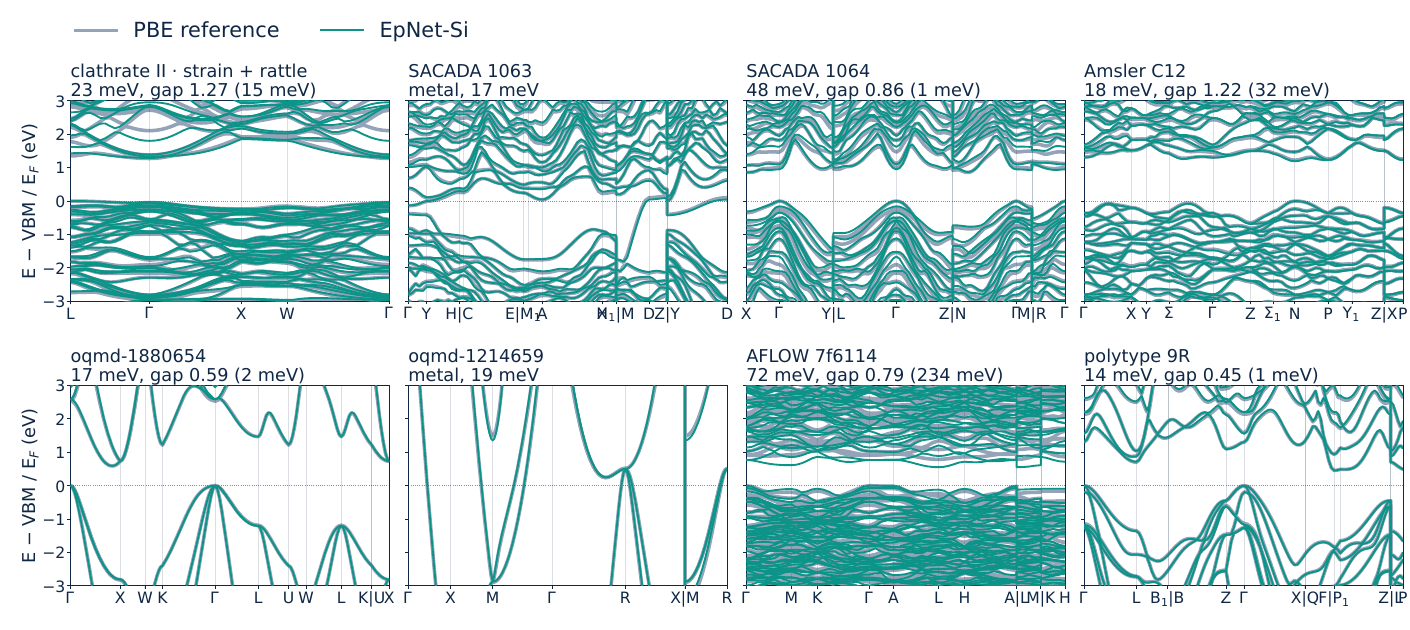}
\caption{\label{fig:new}Unseen structures from each source (clathrate~II, SACADA, Amsler, OQMD, AFLOW, a stacking polytype), picked by name order, not by score. PBE (grey) and EpNet-Si (teal).}
\end{figure}

\begin{figure}[!htbp]
\includegraphics[width=\textwidth]{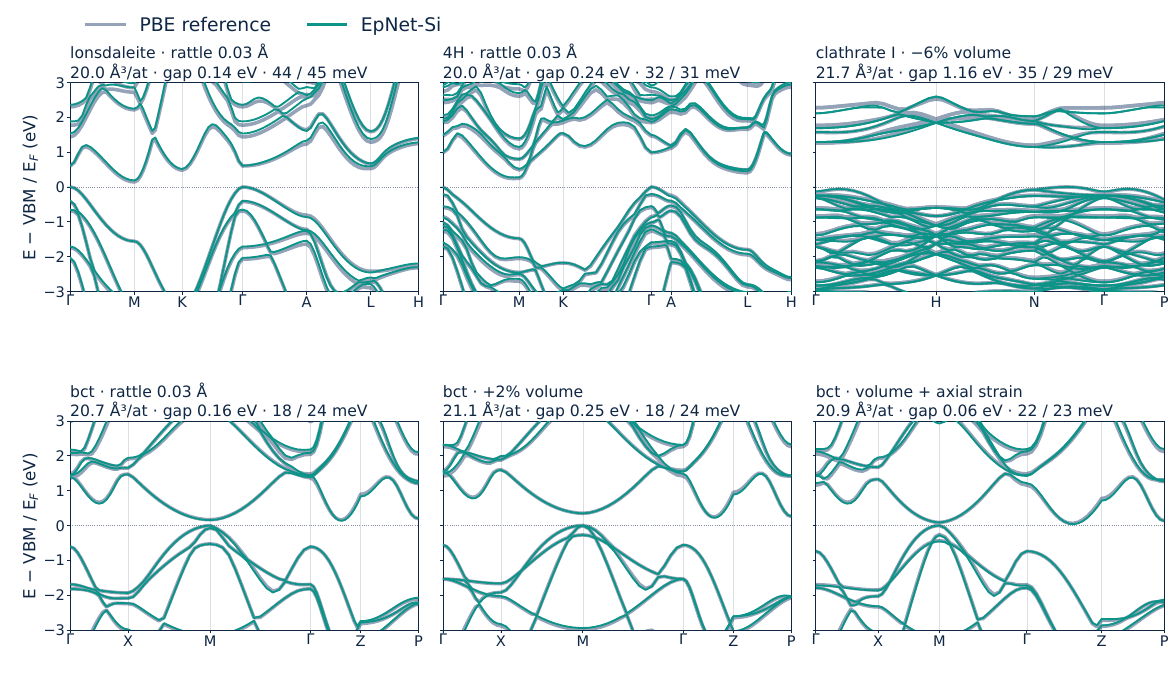}
\caption{\label{fig:knownworst}The six largest gap errors among the 231 distortions of trained structure types: PBE (grey) and EpNet-Si (teal), VBM-aligned.}
\end{figure}

Fig.~\ref{fig:history} shows the training history of EpNet-Si.
Fig.~\ref{fig:known} shows one test cell for each of eight selected trained structure types, and Fig.~\ref{fig:new} shows examples from each source of unseen structures, picked by name order rather than by score.
Fig.~\ref{fig:knownworst} shows the six largest gap errors among the distortions of seen structures; the six worst unseen structures are in Fig.~\ref{fig:worst} of the main text.

\end{document}